\documentclass[aps,prl,twocolumn,amsmath,amssymb,superscriptaddress,longbibliography]{revtex4-1}
\usepackage[english]{babel}
\usepackage{amssymb}
\usepackage{dcolumn}
\usepackage{bm}
\usepackage{graphicx}
\usepackage{amsmath}
\usepackage{graphicx}       % standard LaTeX graphics tool
\usepackage{mathrsfs}                             % when including figure files
\graphicspath{{pict/}{}}
\usepackage{dcolumn}%%%%%new
\usepackage{bm}
\usepackage[pdfstartview=FitH, CJKbookmarks=true, bookmarksnumbered=true, bookmarksopen=true, colorlinks=true, pdfborder=001, citecolor=blue, linkcolor=blue, linktocpage=true] {hyperref}
\begin{document}

 \title{Resolving Vacuum Fluctuations and Nonclassical  States with Cavity-Enhanced Tripartite Interactions}
  \title{Unveiling Vacuum Fluctuations and Nonclassical States with Cavity-Enhanced Tripartite Interactions}
  
\author{Jing Tang}
\affiliation{Guangdong Provincial Key Laboratory of Quantum Metrology and Sensing $\&$ Sc hool of Physics and Astronomy, Sun Yat-Sen University (Zhuhai Campus), Zhuhai 519082, China}

\author{Yuangang Deng}
\email{dengyg3@mail.sysu.edu.cn}
\affiliation{Guangdong Provincial Key Laboratory of Quantum Metrology and Sensing $\&$ Sc hool of Physics and Astronomy, Sun Yat-Sen University (Zhuhai Campus), Zhuhai 519082, China}

\date{\today}
\begin{abstract}
 Enhancing and tailoring light-matter interactions offer remarkable nonlinear resources with wide-ranging applications in various scientific disciplines. In this study, we investigate the construction of strong and deterministic tripartite `beamsplitter' (`squeeze') interactions by utilizing cavity-enhanced nonlinear anti-Stokes (Stokes) scattering within the spin-photon-phonon degrees of freedom. 
We explore the exotic dynamical and steady-state properties associated with the confined motion of a single atom within a high-finesse optical cavity. Notably, we demonstrate the direct extraction of vacuum fluctuations of photons and phonons, which are inherent in Heisenberg's uncertainty principle, without requiring any free parameters. Moreover, our approach enables the realization of high-quality single-quanta sources with large average photon (phonon) occupancies. The underlying physical mechanisms responsible for generating nonclassical quantum emitters are attributed to decay-enhanced single-quanta blockade and the utilization of long-lived motional phonons, resulting in strong nonlinearity. This work unveils significant opportunities for studying hitherto unexplored physical phenomena and provides novel perspectives on fundamental physics dominated by strong tripartite interactions.
\end{abstract} 

\maketitle

{\em Introduction}.---The interface between cavity and quantum matter offer a prominent platform for harnessing the distinctive characteristics of different degrees of freedom, facilitating the engineering of strongly correlated quantum phases~\cite{RevModPhys.85.553,mivehvar2021cavity,PhysRevResearch.5.013002,PhysRevLett.120.123601,leonard2017supersolid,kim2018coherent} and special nonclassical states with wide-ranging applications in quantum technologies~\cite{RevModPhys.73.565,RevModPhys.82.1041,RevModPhys.87.1379,RevModPhys.94.041003}. Remarkable advancements of ground states in low-energy macroscopic mechanical oscillators~\cite{RevModPhys.86.1391,PhysRevLett.99.093902,teufel2011sideband,PhysRevLett.77.4281,PhysRevX.8.021027}, such as optomechanical and center-of-mass motion, has opened up remarkable frontiers for quantum information and sensing~\cite{bennett2000quantum,RevModPhys.91.025005,PhysRevLett.122.030501}, quantum metrology~\cite{PhysRevLett.96.010401,RevModPhys.90.035005,RevModPhys.89.035002}, and studying fundamental physics~\cite{wollman2015quantum,palomaki2013entangling,PhysRevLett.76.1796}. Leveraging their long coherence times, significant advancements have been achieved, ranging from n-quanta bundle states~\cite{munoz2014emitters,PhysRevLett.117.203602,PhysRevLett.124.053601,PhysRevLett.127.073602,deng2021motional} to long-lived phonon-to-optical quantum transducers~\cite{mirhosseini2020superconducting,PhysRevLett.81.5932,forsch2020microwave,bagci2014optical,bochmann2013nanomechanical,PhysRevX.5.031031,PhysRevLett.119.180505}, violation of Bell inequalities~\cite{PhysRevLett.109.230503,arvidsson2020quantum,PhysRevLett.121.220404}, and gravitational wave detection~\cite{PhysRevLett.115.211104,PhysRevLett.118.143601,PhysRevLett.124.221102}. These pioneering explorations primarily focus on the fundamental and ubiquitous physical processes of pairwise coherent light-matter interactions between optical cavity (ideal for quantum networking tasks~\cite{kimble2008quantum}) and long-lived mechanical modes~\cite{RevModPhys.86.1391}. The seminal  interactions are the generalized quantum Rabi model~\cite{PhysRev.49.324,jaynes1963comparison} and nonlinear optomechanical interactions~\cite{RevModPhys.86.1391}.  
  
Meanwhile, the ability to manipulate optomechanical at the single (few) quantum level~\cite{o2010quantum,chu2018creation,arrangoiz2019resolving,PhysRevX.9.021056} and measure high-order phonon correlations~\cite{PhysRevLett.128.183601,riedinger2016non,hong2017hanbury,riedinger2018remote} has opened up new avenues for  experimental realizing novel types of nonlinear interactions, enabling the study of intriguing aspects of quantum mechanics in hybrid quantum systems. Compared to the well-established paradigm of bipartite interactions~\cite{RevModPhys.86.1391,PhysRev.49.324,jaynes1963comparison}, constructing a tripartite interactions involving diverse degrees of freedom holds significant promise for investigating fundamental physics. Fascinating examples include the investigation of high-fidelity tripartite entanglement~\cite{PhysRevLett.92.177903,PhysRevLett.90.167903,PhysRevLett.97.140504,PhysRevA.62.062314,PhysRevLett.82.5385},  novel nonclassical squeezing states~\cite{PhysRevLett.128.143601,PhysRevLett.112.023601,PhysRevLett.116.140402}, and hybrid quantum networks~\cite{PhysRevLett.78.3221,RevModPhys.82.1209,PhysRevX.7.021021} by introducing additional nonlinearities, such as atom and long-live mechanical mode.  Recent theoretical advancements have explored coherent tripartite interactions between a quantum emitter, photon(magnon), and a mechanical oscillator~\cite{PhysRevLett.118.133603,zhou2022synergistic,PhysRevLett.130.073602}. Despite rapid experimental progress, the generation of strong tripartite interactions remains a challenge and the corresponding exotic fundamental quantum phenomena are yet to be fully explored.

In this Letter, we present an experimental scheme to achieve a tunable tripartite interactions using a single alkaline-earth-metal atom deeply trapped in a high-finesse optical cavity. Our architecture, incorporating three distinct degrees of freedom, offers the advantage of of long-lived center-of-mass motion, distinguishing from extensively studied single-atom cavity QEDs~\cite{birnbaum2005photon,mucke2010electromagnetically,PhysRevLett.118.133604}. Specifically, we realize deterministic tripartite spin-photon-phonon `beamsplitter' (`squeeze') interactions, leveraging the cavity-enhanced nonlinear anti-Stokes (Stokes) scattering. By performing the photon and phonon number distributions, we directly resolve the vacuum fluctuations of both the photon and phonon fields. Additionally,  the high-quality single-photon and -photon sources are achieved without requiring single-photon strong coupling.  We demonstrate the nonclassical nature of cavity and motional phonon emissions by analyzing their Wigner functions. Notably, the antibunching amplitudes for both photon and phonon exhibit strongly sensitivity to the dissipation of motional phonon. This result provides a novel method for monitoring the decoherence of hybrid systems via quantum statistics measurement.  In perspective, the strong nonlinearity of the tripartite spin-photon-phonon interactions allows us to clearly encode complex quantum states into the long-lived motional phonon, and pave the way for versatile applications in quantum information and metrology~\cite{lecocq2015resolving,wolf2019motional,ma2021non,RevModPhys.90.045005,pikovski2012probing,PhysRevX.9.041022}.

\begin{figure}[ht]
\centering
\includegraphics[width=0.92\columnwidth]{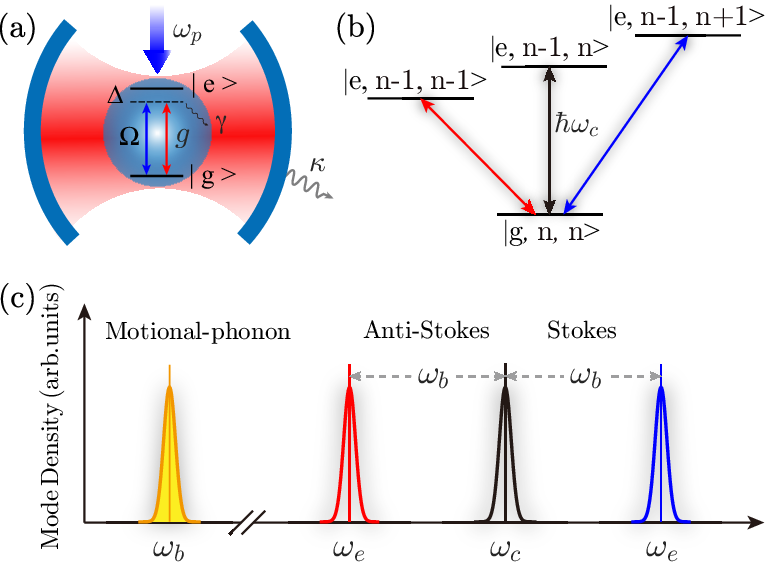}
\caption{(color online).  (a) Schematic diagram of cavity-enhanced tripartite interactions.  (b) Principle of realizing tripartite `beamsplitter'  and `squeeze'  interactions, depicting two distinct processes of red and blue sideband transitions.  (c) Scattering picture of the anti-Stokes (Stokes) process with a photon leading to the annihilation (creation) of a motional phonon.} \label{scheme}
\end{figure}

{\em Model}.--- As illustrated in Fig. ~\ref{scheme}, we consider a single alkaline-earth-metal atom, such as, $^{88}$Sr, confined within a high-finesse cavity and a one-dimensional harmonic trap. The relevant atomic levels consist of a ground state $|g\rangle$ ($^1S_0$) and a long-lived electronic orbital state $|e\rangle$ ($^3P_1$), corresponding to the narrow (7.5-kHz-wide) dipole-forbidden transition. To engineer the dynamic spin-orbit coupling, the atomic transitions $|g\rangle\leftrightarrow |e\rangle$ are resonant coupled by a transverse pump field with Rabi frequency $\Omega$ and quantized standing wave cavity field originated from the collective Bragg scattering. In the Lamb-Dicke regime~\cite{PhysRevA.52.4214},  the tripartite spin-photon-phonon Hamiltonian is given by~\cite{SM} 
\begin{align}
\hat{\cal {H}}/\hbar &= \omega_b\hat{b}^\dag\hat{b} + \Delta_c \hat{a}^\dag\hat{a} + {\frac{\Delta}{2}}\sigma_z +\Omega \hat{\sigma}_{x} \nonumber \\
&+ g (\hat {b}^\dag +\hat {b})(\hat {a}^\dag \hat{\sigma}_{-} + \hat {a}\hat{\sigma}_{+}),
\label{triHam}
\end{align}
where  $\hat{a}$ denotes the annihilation operator of single-mode cavity,  $\hat{b}$  denotes the annihilation operator of bosonic phonon for motional mode with harmonic oscillator frequency $\omega_b$, $\hat{\sigma}_{x,y,z}$ are the Pauli matrices, $\Delta_c$ is the cavity-light detuning, $\Delta$ is the atom-light detuning, and  $g$ is the effective photon-phonon coupling associated with the zero-point fluctuation of the trapped atom oscillator. Apparently, Hamiltonian  (\ref{triHam}) captures the tripartite spin-magnon-optomechanical-type interactions~\cite{zhou2022synergistic,lai2022tripartite,cotrufo2017coherent,hei2023enhanced}, which are fundamental building blocks in quantum optics and photonics community, for example, tripartite entanglement~\cite{PhysRevLett.92.177903,PhysRevLett.90.167903,PhysRevLett.97.140504,PhysRevA.62.062314,PhysRevLett.82.5385}. Interestingly, this interaction allows swapping excitations among an optical mode, quantum emitter, and long-lived motional phonon. Our proposal will facilitate the engineering of hybrid spin-photon-phonon quantum transducers and bridge the different research areas of science. 

To capture the underlying of physics, one can obtain a tripartite `beamsplitter' interactions~\cite{SM}  
\begin{align}\label{triBS}
{\cal\hat{H}}_{\rm{B}}/\hbar
&={\frac{\Delta}{2}}\sigma_z +\Omega \hat{\sigma}_{x} + g (\hat {a}^\dag \hat {b} \hat{\sigma}_{-} + \hat {b}^\dag\hat {a}\hat{\sigma}_{+}),
\end{align}
under the rotating-wave approximation (RWA) with $g/\omega_b\ll 1$ and red-sideband resonance $\Delta_c/\omega_b=1$. This nonlinear anti-Stokes process denotes an annihilation of a motional phonon up converting a pump photon into a cavity photon, corresponding to the coherent exchange among quantum emitter, cavity, and phonon fields.  For firstly neglecting the weak pump field ($\Omega=0$), the system exhibits a ${\cal U}(1)$ symmetry satisfying the commutation relation, $[{\cal R}_{\theta}, {\cal\hat{H}}_{\rm{B}}]=0$, where  ${\cal R}_{\theta} =\exp[-i\theta(2\hat{a}^\dag\hat{a}+\hat{b}^\dag\hat{b} +\sigma_z/2)]$ is the action of the operator. The occurrence of ${\cal U}(1)$ symmetry breaking corresponds to single-atom superradiance~\cite{kim2018coherent}.

Under the RWA ($g/\omega_b\ll 1$) and blue-sideband resonance ($\Delta_c/\omega_b\approx -1$), Hamiltonian (\ref{triHam}) will reduce to tripartite `squeeze'  interactions in a rotating reference frame at the frequency $\omega'$ ($\approx \omega_b$) 
\begin{align}\label{triSque}
{\cal\hat{H}}_{\rm{S}}/\hbar
&=  \delta_b\hat{b}^\dag\hat{b} + \delta_a \hat{a}^\dag\hat{a} +  {\frac{\delta_a-\delta}{2}}\sigma_z +\Omega \hat{\sigma}_{x}  \nonumber \\
&+ g(\hat {a}^\dag \hat {b}^\dag \hat{\sigma}_{-} + \hat {b}\hat {a}\hat{\sigma}_{+}),
\end{align}
where $\delta_b=\omega_b-\omega'$, $\delta_a=\Delta_c+\omega'$, and $\delta=\delta_a-\Delta$ is the detuning of the cavity emission frequency from the atomic resonance. The Hamiltonian (\ref{triSque}) captures the cavity-enhanced Stokes scattering, representing a deterministic parametric down-conversion process with respect to the pump photon down-converted into entanglement of cavity and motional phonon. For $\Omega=0$, these tripartite interactions also exhibit another ${\cal U}(1)$ symmetry characterized by the operator, ${\cal R}_{\theta} =\exp[-i\theta(\hat{a}^\dag\hat{a}+\hat{b}^\dag\hat{b} +\sigma_z)]$, satisfying ${\cal R}_{\theta}^\dag(\hat{a},\hat{b},\sigma_-,\sigma_+){\cal R}_{\theta} =(\hat{a} e^{-i\theta},\hat{b} e^{-i\theta},\sigma_-e^{-i2\theta},\sigma_+e^{i2\theta})$. For single-atom superradiance, the strong nonclassical correlated photon-phonon pairs in the single-quanta level are expected since the emitted photon and phonon are simultaneously created or destroyed. 

In contrast to the generation of correlated single-photon pairs via parametric down-conversion process in bimodal cavities~\cite{PhysRevA.105.043711},  our hybridization system highlights the direct entering the strong-coupling regime with the tripartite single-atom cooperativity ${\cal C}=g^3/(\gamma\kappa_a\kappa_b)\gg 1$, as both motional phonon and atomic excited state possess long lifetimes.  Here, $\gamma$ is atomic spontaneous emission rate of the excited state, $\kappa_a$ and $\kappa_b$ are the decay rates of the photon and phonon, respectively. In our numerical simulation,  we adopt
$\gamma=7.5~{\rm kHz}~(2\pi)$~\cite{RevModPhys.87.637,PhysRevX.6.011025,PhysRevLett.118.263601},  $\kappa_b/\gamma=1$, $\kappa_a/\gamma=10 $~\cite{PhysRevX.6.011025,PhysRevLett.118.263601,wolke2012cavity}, and $\Omega/\gamma=0.2 \kappa_a$. The single atom-cavity coupling is $g/\gamma=40$,  which has been demonstrated in the state-of-the-art cavity QEDs experiments~\cite{PhysRevLett.118.133604,leonard2017supersolid,leonard2017monitoring}. The independent controlled parameters are $\omega_b$ and $\Delta_c$.

\begin{figure}[ht]       
\includegraphics[width=0.95\columnwidth]{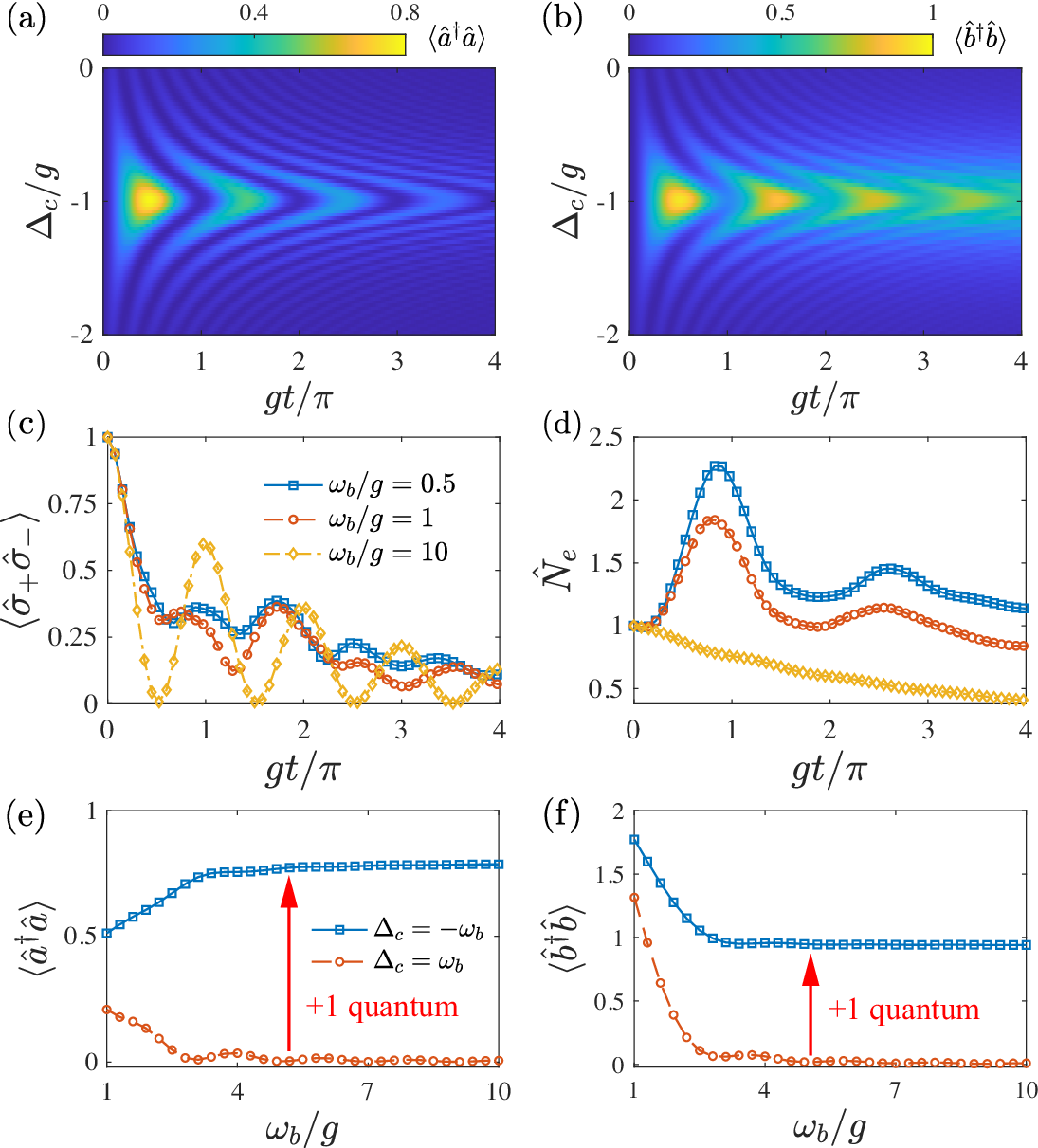}
\caption{(color online). (a)-(b) Photon and phonon occupations response as a function of $t$ and $\Delta_c$ with  $\omega_b/g=10$. (c)-(d) Evolution of  $\langle \hat{\sigma}_+\hat{\sigma}_-\rangle$ and $ \hat{N}_e$ for different $\omega_b$ with $\Delta_c/\omega_b=-1$.   (e)-(f) $\omega_b$ dependence of  $\langle \hat{a}^\dag \hat{a}\rangle$ and $\langle \hat{b}^\dag \hat{b}\rangle$  for different $\Delta_c$ at $gt/\pi=0.5$. The other parameter is $\Delta=0$. } \label{dynamics}
\end{figure}

{\em Results}.---To investigate the out-of-equilibrium dynamics, we analyze the quantum statistics of photon and phonon emissions by solving the master equation considering the complete dissipations of the system~\cite{SM}. Figure \ref{dynamics} presents the numerical results depicting the time-dependent occupations and correlation evolutions. The single atom is initially prepared in the excited state, while the cavity and phonon fields are in the vacuum state. Figures \ref{dynamics}(a) and \ref{dynamics}(b) illustrate the population of photons and motional phonons in $t\Delta_c$ parameter plane.  For the blue sideband focus $\Delta_c/\omega_b\approx -1$,  the periodic oscillations of the quantized fields are dominated by the tripartite `squeeze' interactions described by Eq.~(\ref{triSque}). These oscillations correspond to the vacuum Rabi frequency $\sqrt{g^2+(\omega_b+\Delta_c)^2/4}$. With time evolution, the Rabi oscillation exhibits a clear pattern of complete collapse and posterior revival with a gradually decreasing amplitude. The difference in amplitude between the photon and phonon oscillations arises due to the different decay rates $\kappa_a$ and $\kappa_b$.  In Fig.~\ref{dynamics}(c), we perform the evolution of spin excitation  $\langle \hat{\sigma}_+\hat{\sigma}_-\rangle$  for different phonon frequencies $\omega_b$. In contrast to the Rabi oscillation observed for  $\omega_b/g\gg1$, the average spin excitation $\langle \hat{\sigma}_+\hat{\sigma}_-\rangle$ in the strong coupling regime ($\omega_b/g\leqslant1$) does not display a complete collapse and revival. Moreover, the total excitation $ \hat{N}_e=\langle \hat{\sigma}_+\hat{\sigma}_-\rangle + [\langle \hat{a}^\dag \hat{a}\rangle +\langle \hat{b}^\dag \hat{b}\rangle]/2$ clearly exceed 1 [Fig.~\ref{dynamics}(d)]. This departure clearly demonstrates that the out-of-equilibrium dynamics go beyond the tripartite `squeeze' interactions, as RWA is breaks down for small phonon frequency ($\omega_b/g\leqslant1$). Indeed, the total excitation number remains  conserved satisfying  $[\hat{N}_e, {\cal\hat{H}}_{\rm{S}}]=0$ when the system dissipations are neglected. 

Figures \ref{dynamics}(e) and \ref{dynamics}(f)  show the photon and phonon occupations  as a function of $\omega_b$  for different values of $\Delta_c$. Clearly,  $\langle \hat{a}^\dag \hat{a}\rangle$ and $\langle \hat{b}^\dag \hat{b}\rangle$ both decrease, eventually saturating at zero at the red sideband ($\Delta_c/\omega_b= 1$). While for the blue sideband ($\Delta_c/\omega_b= -1$),  the average number of photon (phonon) is gradually increasing (decreasing), reaching saturation values for  $\omega_b/g \gg1$. Analogous to bipartite Rabi model~\cite{PhysRevX.8.021027}, the observed $\langle \hat{b}^\dag \hat{b}\rangle>1$ is the hallmark for entering the strong coupling regime in tripartite model. Remarkably, the difference in cavity (phonon) occupancies between the blue and red sidebands is approximately equal to one, extracted by full numerical simulations that take into account the finite linewidth of each mode. This result unambiguously demonstrates the vacuum fluctuations of the cavity and motional phonon originally Heisenberg's uncertainty principle. 

To acquire more physical insight, we can study the dynamics in the Schr\"{o}dinger picture, describing the time evolution of photon and phonon amplitudes excluding the dissipations of system. The time-dependent states after the nonlinear Stokes and anti-Stokes processes read~\cite{SM}
\begin{align}
|\psi(t)\rangle_S  &=\cos\theta_S|n,n,e\rangle -i \sin\theta_S |n+1,n+1,g\rangle, \nonumber \\
|\psi(t)\rangle_B  &=\cos\theta_B|n,n,e\rangle -i \sin\theta_B |n+1,n-1,g\rangle,
\end{align}
where $n$ represents the initial excitation of quanta in the quantized fields, $\theta_S=g(n+1)t$ and $\theta_B=g\sqrt{n(n+1)}t$ are accumulated phases after tripartite `squeeze'  and `beamsplitter' interactions, respectively. It is evident that the average photon and phonon occupancies satisfy $\langle \hat{a}^\dag \hat{a}\rangle= \langle \hat{b}^\dag \hat{b}\rangle=1$ at $t=\pi/2g$ even when the quantized fields are prepared in vacuum ($n=0$). This parametric amplification proceeds by cavity-enhanced Stokes scattering. Conversely, the cooling process leads to $\langle \hat{a}^\dag \hat{a}\rangle= \langle \hat{b}^\dag \hat{b}\rangle=0$. Notably, the observed vacuum fluctuations reveal a striking signature of quantum nature, associated with the commutation relation, $[\hat{o},\hat{o}^\dag]=1$. The advantage of the tripartite device is that vacuum fluctuations can be directly obtained without relying on free parameters. This is in contrast to pioneering experiment in bipartite optomechanics, where vacuum fluctuations were resolved by normalizing the ratio of the final cavity displacement to the initial mechanical displacement~\cite{lecocq2015resolving}.

\begin{figure}[ht]
\includegraphics[width=0.95\columnwidth]{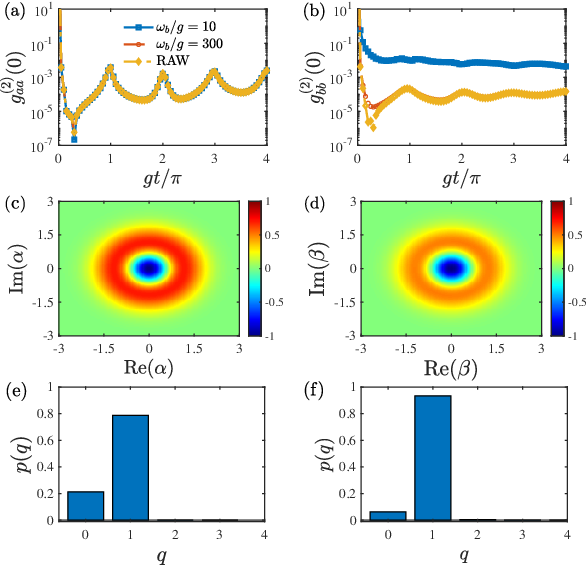}
\caption{(color online). (a)-(b) The evolution of  $g^{(2)}_{aa}(0)$ and $g^{(2)}_{bb}(0)$ for different  $\omega_b$ with $\Delta_c/\omega_b=-1$.  (c)-(d) Wigner functions of the photon and phonon at $gt/\pi=0.5$ and $\omega_b/g=10$.  (g)-(h) The corresponding photon- and phonon-number distributions $p(q)$. The remaining parameter is $\Delta=0$. } \label{wigner}
\end{figure}

To gain further insight into the quantum statistics, we plot second-order correlation function $g_{aa}^{(2)}(0)$ and $g_{bb}^{(2)}(0)$  for different $\omega_b$ in the weak coupling regime, as displayed in Figs.~\ref{wigner}(a) and \ref{wigner}(b).  We find that the value of $g_{aa}^{(2)}(0)$ for the cavity is independent on large phonon frequency. Remarkably, we achieve the strong instantaneous photon blockade with $g_{aa}^{(2)}(0)=2.5\times10^{-5}$ and an occupation of $\langle \hat{a}^\dag \hat{a}\rangle=0.79$ at $gt/\pi=0.5$, which demonstrates the potential for a high-quality single-photon source even with weak single atom-cavity coupling  ($g/\kappa_{a}=4$). Additionally, we check that both photon and phonon occupations are robust against variations in the large value of $\omega_b$~\cite{SM}. The intuitive understanding is that RWA is valid when $\omega_b/g>10$. Interestingly, the second-order correlation for the motional phonon exhibits high sensitivity to  $\omega_b$. The phonon antibunching amplitude increases rapidly with $\omega_b$. For RWA (equivalent to $\omega_b/g\sim\infty$), the high-quality single-phonon source is realized with $g_{bb}^{(2)}(0)=2.3\times10^{-5}$ and occupation $\langle \hat{b}^\dag \hat{b}\rangle=0.94$ at $gt/\pi=0.5$. These results indicate that $g_{bb}^{(2)}(0)$ can not be exclusively determined by the tripartite `squeeze'  Hamiltonian even for $\omega_b/g\gg1$, when far-resonance high-frequency terms $g (e^{-i2\omega_bt}\hat {a}^\dag \hat {b} \hat{\sigma}_{-} + {\rm H.c.})$ are ignored. We note that the quantum statistics for the cavity photon and motional phonon can be experimentally measured by utilizing  Hanbury, Brown, and Twiss interferometer~\cite{birnbaum2005photon} and spin state-resolved projective measurements~\cite{PhysRevX.8.021027,wolf2019motional,PhysRevLett.128.183601}.  

\begin{figure}[ht]
\includegraphics[width=0.95\columnwidth]{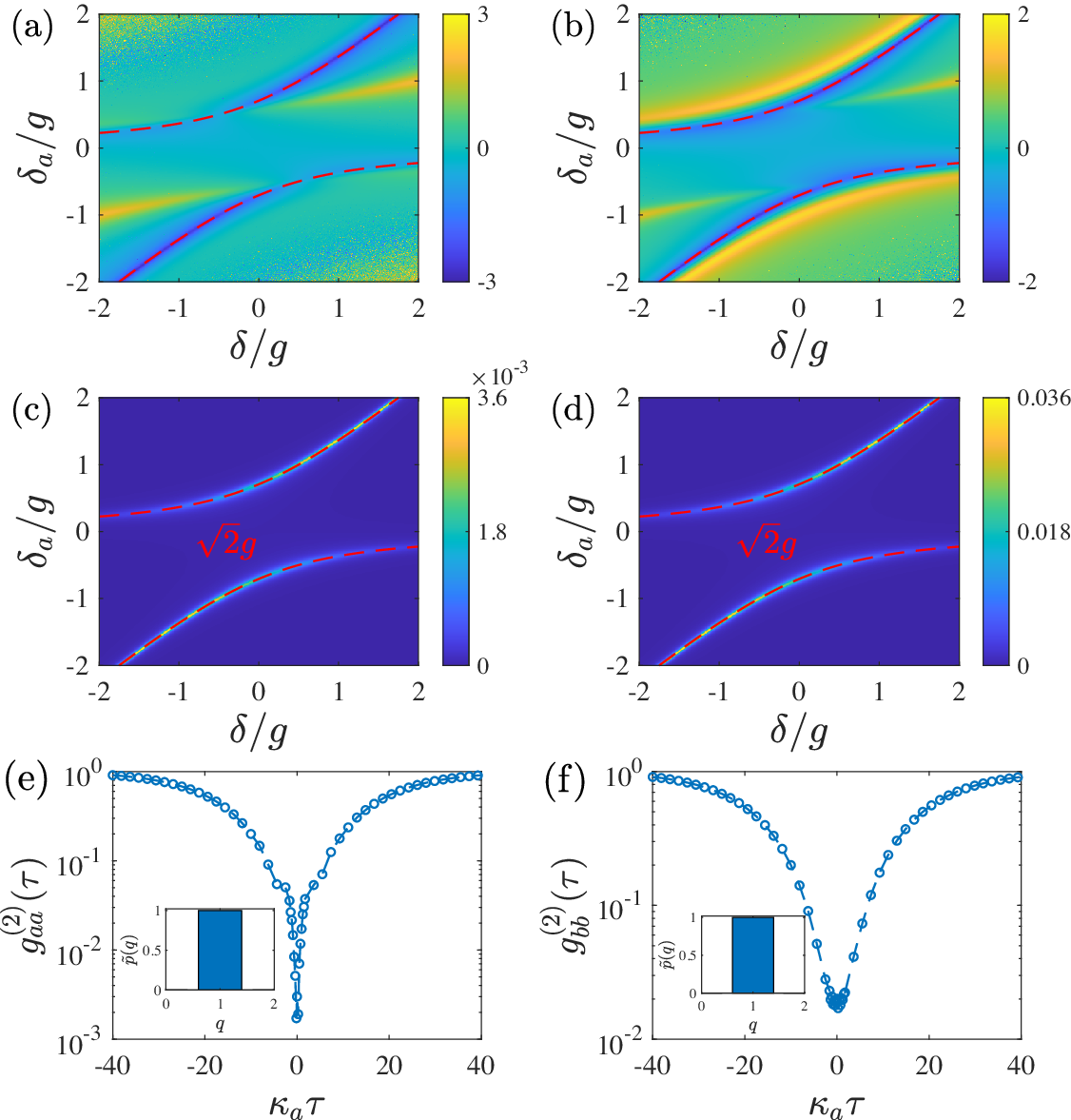}
\caption{(color online). (a)-(b) show the values of log$[g^{(2)}_{aa}(0)]$ and  log$[g^{(2)}_{bb}(0)]$ on $\delta\delta_a$ parameter plane. (c)-(d) The corresponding steady-state population $n_a^{(s)}$ and $n_b^{(s)}$. The red dashed lines denote the analytic vacuum Rabi splittings in the upper and lower branches. (e)-(f) $\tau$ dependence of $g^{(2)}_{aa}(\tau)$ and $g^{(2)}_{bb}(\tau)$ with $\delta_a/g=(1+\sqrt{3})/2$ and $\delta/g=1$. Inset displays the steady-state distribution $\tilde p(q)$. The other parameters are $\delta_a/\delta_b=1$ and $\omega'/g=100$ } \label{NMsplitting}
\end{figure}

The concept of quasiprobability distribution, such as Wigner function,  leads to tremendously successful in modern quantum physics. In order to extract the quantum nature of the realized single-quanta states, we calculate Wigner functions $W(\alpha)$ and $W(\beta)$ in phase-space amplitudes $\alpha$ and $\beta$~\cite{PhysRevA.15.449}.  The calculated results, shown in  Figs.~\ref{wigner}(c) and \ref{wigner}(d), reveal the presence of negative values in Wigner function, unambiguously demonstrating the nonclassical nature of cavity and motional phonon. Although experimentally certifying nonclassicality remains a challenging task,  the full tomographic reconstruction of Wigner function and related phase-space distributions can be extracted by performing correlation measurement techniques of cavity~\cite{PhysRevLett.124.133601,PhysRevLett.126.023605} and phonon number distribution in mechanical resonator~\cite{chu2018creation}. Figures \ref{wigner}(e) and \ref{wigner}(f) depict the photon- and phonon-number distributions $p(q)={\rm Tr}(|q\rangle\langle q| \rho)$, which provide a quantitative measure of the quality of single-quanta emissions. Clearly, the probability of multi-quanta excitations ($n\geqslant 2$) roughly zero reveals the single-quanta nature for photons and phonons, which agrees well with second-order correlation function $g_{aa}^{(2)}(0)$ and $g_{bb}^{(2)}(0)$ presented in Figs.~\ref{wigner}(a) and \ref{wigner}(b). Thus our system serves as the high-quality single-photon and single-phonon source with a clear signature of nonclassicality. 

Besides out-of-equilibrium dynamics, we further explore the steady-state properties of the tripartite `squeeze'  interactions (Stokes process) with $\omega'/g=100$. We emphasize that both photon and phonon emissions are independent of the initial state of the atomic field in a realistic scenario. Figures \ref{NMsplitting}(a) and  \ref{NMsplitting}(b) display the second-order correlation functions in the $\delta\delta_a$ parameter plane. It is shown that the values of $g^{(2)}_{aa}(0)$ for photons and $g^{(2)}_{bb}(0)$ for motional phonons vary over a wide range. Remarkably,  we achieve significant photon and phonon blockades with $g^{(2)}_{aa}(0)\sim10^{-3}$ and $g^{(2)}_{bb}(0)\sim10^{-2}$, which represent the hallmark of strong photon and phonon antibunching.  As expected, strong antibunching occurs at the single-quanta resonance of upper and lower branches, which is in excellent agreement with the analytic vacuum Rabi splittings satisfying $\delta^{(\pm)} _a=[\delta \pm \sqrt{2g^{2} +\delta^2}]/2$ derived from the energy spectrum (red dashed lines)~\cite{SM}.  We find that the energy spectrum anharmonicity is enhanced by tuning $\delta$, resulting in significantly amplified antibunching amplitudes for the quantized fields. The distinct avoided crossing features for photon and phonon emissions are clear signatures of cavity-atom and phonon-atom normal-mode splittings, as depicted in Figs. \ref{NMsplitting}(c) and  \ref{NMsplitting}(d), corresponding to a minimum value of $\sqrt{2}g$ for both cavity and motional phonon modes. 

To further analyze the  quantum coherence maintained within dissipations, we plot interval $\tau$ dependence of $g^{(2)}_{aa}(\tau)$ and $g^{(2)}_{bb}(\tau)$, as shown in Figs.~\ref{NMsplitting}(e) and \ref{NMsplitting}(f). Interestingly, the decay times for both photon and phonon are obviously longer than the typical time scale of $\kappa_a^{-1}$, thanks to the advantageous properties of long-lived motional phonons in the tripartite spin-photon-phonon system. The single-quanta blockade nature is further confirmed through the steady-state distribution $\tilde p(q)\equiv qp(q)/n^{(s)}$, which characterizes the fraction of $q$-quanta states among the total excitations. Remarkably, we observe strong photon- and phonon-blockade with the typical probability of multiquanta excitations ($n>2$) being below $0.001\%$ and $0.07\%$, respectively. We verify that a large decay rate corresponds to a strong antibunching amplitude and a small occupation number of emissions~\cite{SM}. This distinct generation of strong single-quanta blockade differs from the study of single-photon pairs in bimodal cavities by utilizing the quantum interference suppressed two-photon excitations~\cite{PhysRevA.105.043711}. The mechanism of decay-enhanced strong single-quanta blockade could facilitate the experimental feasibility for engineering special nonclassical states beyond the limits of strong atom-cavity coupling.

{\em Conclusion}.---We propose a novel architecture incorporates a single atom with its long-lived motional degree of freedom within an optical cavity leveraging state-of-the-art experimental techniques. By realizing two distinct types of deterministic tripartite spin-photon-phonon interactions, we have realized an interesting nonlinear resource for generating strong nonclassical quantum emitters. Compared to the experimentally resolving vacuum fluctuations in bipartite optomechanical system~\cite{lecocq2015resolving}, our tripartite system enables directly extract vacuum fluctuations inherently origin Heisenberg's uncertainty principle without free parameters. Moreover, the high-quality single-quanta emissions are demonstrated that surpass the limits of strong single atom-cavity coupling. This achievement is attributed to the mechanism of decay-enhanced single-quanta blockade. By combining additional strong nonlinearity and exquisite manipulation of individual photon and long-lived mechanical mode, our results offer a platform for integrating the advantageous features of atom, optical cavity, and low-energy center-of-mass motion. This integrated approach fosters tremendous progress for exploring the tripartite entanglement~\cite{PhysRevLett.92.177903,PhysRevLett.90.167903,PhysRevLett.97.140504,PhysRevA.62.062314,PhysRevLett.82.5385}, decoherence mechanism~\cite{arndt2014testing}, and distributed quantum information processing~\cite{PhysRevA.89.022317,fitzsimons2017private,RevModPhys.82.665,RevModPhys.94.041003}.

\section*{ACKNOWLEDGMENTS}
This work was supported by the NSFC (Grant No. 12274473 and Grant No. 12135018) and National Key R$\&$D Program of China (Grant No. 2018YFA0307500).

%\bibliographystyle{apsrev4-1.bst}
%\bibliography{ref_QF}

\begin{thebibliography}{99}%
\makeatletter
\providecommand \@ifxundefined [1]{%
 \@ifx{#1\undefined}
}%
\providecommand \@ifnum [1]{%
 \ifnum #1\expandafter \@firstoftwo
 \else \expandafter \@secondoftwo
 \fi
}%
\providecommand \@ifx [1]{%
 \ifx #1\expandafter \@firstoftwo
 \else \expandafter \@secondoftwo
 \fi
}%
\providecommand \natexlab [1]{#1}%
\providecommand \enquote  [1]{``#1''}%
\providecommand \bibnamefont  [1]{#1}%
\providecommand \bibfnamefont [1]{#1}%
\providecommand \citenamefont [1]{#1}%
\providecommand \href@noop [0]{\@secondoftwo}%
\providecommand \href [0]{\begingroup \@sanitize@url \@href}%
\providecommand \@href[1]{\@@startlink{#1}\@@href}%
\providecommand \@@href[1]{\endgroup#1\@@endlink}%
\providecommand \@sanitize@url [0]{\catcode `\\12\catcode `\$12\catcode
  `\&12\catcode `\#12\catcode `\^12\catcode `\_12\catcode `\%12\relax}%
\providecommand \@@startlink[1]{}%
\providecommand \@@endlink[0]{}%
\providecommand \url  [0]{\begingroup\@sanitize@url \@url }%
\providecommand \@url [1]{\endgroup\@href {#1}{\urlprefix }}%
\providecommand \urlprefix  [0]{URL }%
\providecommand \Eprint [0]{\href }%
\providecommand \doibase [0]{http://dx.doi.org/}%
\providecommand \selectlanguage [0]{\@gobble}%
\providecommand \bibinfo  [0]{\@secondoftwo}%
\providecommand \bibfield  [0]{\@secondoftwo}%
\providecommand \translation [1]{[#1]}%
\providecommand \BibitemOpen [0]{}%
\providecommand \bibitemStop [0]{}%
\providecommand \bibitemNoStop [0]{.\EOS\space}%
\providecommand \EOS [0]{\spacefactor3000\relax}%
\providecommand \BibitemShut  [1]{\csname bibitem#1\endcsname}%
\let\auto@bib@innerbib\@empty
%</preamble>
\bibitem [{\citenamefont {Ritsch}\ \emph {et~al.}(2013)\citenamefont {Ritsch},
  \citenamefont {Domokos}, \citenamefont {Brennecke},\ and\ \citenamefont
  {Esslinger}}]{RevModPhys.85.553}%
  \BibitemOpen
  \bibfield  {author} {\bibinfo {author} {\bibfnamefont {H.}~\bibnamefont
  {Ritsch}}, \bibinfo {author} {\bibfnamefont {P.}~\bibnamefont {Domokos}},
  \bibinfo {author} {\bibfnamefont {F.}~\bibnamefont {Brennecke}}, \ and\
  \bibinfo {author} {\bibfnamefont {T.}~\bibnamefont {Esslinger}},\ }\href
  {\doibase 10.1103/RevModPhys.85.553} {\bibfield  {journal} {\bibinfo
  {journal} {Rev. Mod. Phys.}\ }\textbf {\bibinfo {volume} {85}},\ \bibinfo
  {pages} {553} (\bibinfo {year} {2013})}\BibitemShut {NoStop}%
\bibitem [{\citenamefont {Mivehvar}\ \emph {et~al.}(2021)\citenamefont
  {Mivehvar}, \citenamefont {Piazza}, \citenamefont {Donner},\ and\
  \citenamefont {Ritsch}}]{mivehvar2021cavity}%
  \BibitemOpen
  \bibfield  {author} {\bibinfo {author} {\bibfnamefont {F.}~\bibnamefont
  {Mivehvar}}, \bibinfo {author} {\bibfnamefont {F.}~\bibnamefont {Piazza}},
  \bibinfo {author} {\bibfnamefont {T.}~\bibnamefont {Donner}}, \ and\ \bibinfo
  {author} {\bibfnamefont {H.}~\bibnamefont {Ritsch}},\ }\href@noop {}
  {\bibfield  {journal} {\bibinfo  {journal} {Advances in Physics}\ }\textbf
  {\bibinfo {volume} {70}},\ \bibinfo {pages} {1} (\bibinfo {year}
  {2021})}\BibitemShut {NoStop}%
\bibitem [{\citenamefont {Deng}\ and\ \citenamefont
  {Yi}(2023)}]{PhysRevResearch.5.013002}%
  \BibitemOpen
  \bibfield  {author} {\bibinfo {author} {\bibfnamefont {Y.}~\bibnamefont
  {Deng}}\ and\ \bibinfo {author} {\bibfnamefont {S.}~\bibnamefont {Yi}},\
  }\href {\doibase 10.1103/PhysRevResearch.5.013002} {\bibfield  {journal}
  {\bibinfo  {journal} {Phys. Rev. Res.}\ }\textbf {\bibinfo {volume} {5}},\
  \bibinfo {pages} {013002} (\bibinfo {year} {2023})}\BibitemShut {NoStop}%
\bibitem [{\citenamefont {Mivehvar}\ \emph {et~al.}(2018)\citenamefont
  {Mivehvar}, \citenamefont {Ostermann}, \citenamefont {Piazza},\ and\
  \citenamefont {Ritsch}}]{PhysRevLett.120.123601}%
  \BibitemOpen
  \bibfield  {author} {\bibinfo {author} {\bibfnamefont {F.}~\bibnamefont
  {Mivehvar}}, \bibinfo {author} {\bibfnamefont {S.}~\bibnamefont {Ostermann}},
  \bibinfo {author} {\bibfnamefont {F.}~\bibnamefont {Piazza}}, \ and\ \bibinfo
  {author} {\bibfnamefont {H.}~\bibnamefont {Ritsch}},\ }\href {\doibase
  10.1103/PhysRevLett.120.123601} {\bibfield  {journal} {\bibinfo  {journal}
  {Phys. Rev. Lett.}\ }\textbf {\bibinfo {volume} {120}},\ \bibinfo {pages}
  {123601} (\bibinfo {year} {2018})}\BibitemShut {NoStop}%
\bibitem [{\citenamefont {L{\'e}onard}\ \emph
  {et~al.}(2017{\natexlab{a}})\citenamefont {L{\'e}onard}, \citenamefont
  {Morales}, \citenamefont {Zupancic}, \citenamefont {Esslinger},\ and\
  \citenamefont {Donner}}]{leonard2017supersolid}%
  \BibitemOpen
  \bibfield  {author} {\bibinfo {author} {\bibfnamefont {J.}~\bibnamefont
  {L{\'e}onard}}, \bibinfo {author} {\bibfnamefont {A.}~\bibnamefont
  {Morales}}, \bibinfo {author} {\bibfnamefont {P.}~\bibnamefont {Zupancic}},
  \bibinfo {author} {\bibfnamefont {T.}~\bibnamefont {Esslinger}}, \ and\
  \bibinfo {author} {\bibfnamefont {T.}~\bibnamefont {Donner}},\ }\href@noop {}
  {\bibfield  {journal} {\bibinfo  {journal} {Nature}\ }\textbf {\bibinfo
  {volume} {543}},\ \bibinfo {pages} {87} (\bibinfo {year}
  {2017}{\natexlab{a}})}\BibitemShut {NoStop}%
\bibitem [{\citenamefont {Kim}\ \emph {et~al.}(2018)\citenamefont {Kim},
  \citenamefont {Yang}, \citenamefont {Oh},\ and\ \citenamefont
  {An}}]{kim2018coherent}%
  \BibitemOpen
  \bibfield  {author} {\bibinfo {author} {\bibfnamefont {J.}~\bibnamefont
  {Kim}}, \bibinfo {author} {\bibfnamefont {D.}~\bibnamefont {Yang}}, \bibinfo
  {author} {\bibfnamefont {S.-h.}\ \bibnamefont {Oh}}, \ and\ \bibinfo {author}
  {\bibfnamefont {K.}~\bibnamefont {An}},\ }\href@noop {} {\bibfield  {journal}
  {\bibinfo  {journal} {Science}\ }\textbf {\bibinfo {volume} {359}},\ \bibinfo
  {pages} {662} (\bibinfo {year} {2018})}\BibitemShut {NoStop}%
\bibitem [{\citenamefont {Raimond}\ \emph {et~al.}(2001)\citenamefont
  {Raimond}, \citenamefont {Brune},\ and\ \citenamefont
  {Haroche}}]{RevModPhys.73.565}%
  \BibitemOpen
  \bibfield  {author} {\bibinfo {author} {\bibfnamefont {J.~M.}\ \bibnamefont
  {Raimond}}, \bibinfo {author} {\bibfnamefont {M.}~\bibnamefont {Brune}}, \
  and\ \bibinfo {author} {\bibfnamefont {S.}~\bibnamefont {Haroche}},\ }\href
  {\doibase 10.1103/RevModPhys.73.565} {\bibfield  {journal} {\bibinfo
  {journal} {Rev. Mod. Phys.}\ }\textbf {\bibinfo {volume} {73}},\ \bibinfo
  {pages} {565} (\bibinfo {year} {2001})}\BibitemShut {NoStop}%
\bibitem [{\citenamefont {Hammerer}\ \emph {et~al.}(2010)\citenamefont
  {Hammerer}, \citenamefont {S\o{}rensen},\ and\ \citenamefont
  {Polzik}}]{RevModPhys.82.1041}%
  \BibitemOpen
  \bibfield  {author} {\bibinfo {author} {\bibfnamefont {K.}~\bibnamefont
  {Hammerer}}, \bibinfo {author} {\bibfnamefont {A.~S.}\ \bibnamefont
  {S\o{}rensen}}, \ and\ \bibinfo {author} {\bibfnamefont {E.~S.}\ \bibnamefont
  {Polzik}},\ }\href {\doibase 10.1103/RevModPhys.82.1041} {\bibfield
  {journal} {\bibinfo  {journal} {Rev. Mod. Phys.}\ }\textbf {\bibinfo {volume}
  {82}},\ \bibinfo {pages} {1041} (\bibinfo {year} {2010})}\BibitemShut
  {NoStop}%
\bibitem [{\citenamefont {Reiserer}\ and\ \citenamefont
  {Rempe}(2015)}]{RevModPhys.87.1379}%
  \BibitemOpen
  \bibfield  {author} {\bibinfo {author} {\bibfnamefont {A.}~\bibnamefont
  {Reiserer}}\ and\ \bibinfo {author} {\bibfnamefont {G.}~\bibnamefont
  {Rempe}},\ }\href {\doibase 10.1103/RevModPhys.87.1379} {\bibfield  {journal}
  {\bibinfo  {journal} {Rev. Mod. Phys.}\ }\textbf {\bibinfo {volume} {87}},\
  \bibinfo {pages} {1379} (\bibinfo {year} {2015})}\BibitemShut {NoStop}%
\bibitem [{\citenamefont {Reiserer}(2022)}]{RevModPhys.94.041003}%
  \BibitemOpen
  \bibfield  {author} {\bibinfo {author} {\bibfnamefont {A.}~\bibnamefont
  {Reiserer}},\ }\href {\doibase 10.1103/RevModPhys.94.041003} {\bibfield
  {journal} {\bibinfo  {journal} {Rev. Mod. Phys.}\ }\textbf {\bibinfo {volume}
  {94}},\ \bibinfo {pages} {041003} (\bibinfo {year} {2022})}\BibitemShut
  {NoStop}%
\bibitem [{\citenamefont {Aspelmeyer}\ \emph {et~al.}(2014)\citenamefont
  {Aspelmeyer}, \citenamefont {Kippenberg},\ and\ \citenamefont
  {Marquardt}}]{RevModPhys.86.1391}%
  \BibitemOpen
  \bibfield  {author} {\bibinfo {author} {\bibfnamefont {M.}~\bibnamefont
  {Aspelmeyer}}, \bibinfo {author} {\bibfnamefont {T.~J.}\ \bibnamefont
  {Kippenberg}}, \ and\ \bibinfo {author} {\bibfnamefont {F.}~\bibnamefont
  {Marquardt}},\ }\href {\doibase 10.1103/RevModPhys.86.1391} {\bibfield
  {journal} {\bibinfo  {journal} {Rev. Mod. Phys.}\ }\textbf {\bibinfo {volume}
  {86}},\ \bibinfo {pages} {1391} (\bibinfo {year} {2014})}\BibitemShut
  {NoStop}%
\bibitem [{\citenamefont {Marquardt}\ \emph {et~al.}(2007)\citenamefont
  {Marquardt}, \citenamefont {Chen}, \citenamefont {Clerk},\ and\ \citenamefont
  {Girvin}}]{PhysRevLett.99.093902}%
  \BibitemOpen
  \bibfield  {author} {\bibinfo {author} {\bibfnamefont {F.}~\bibnamefont
  {Marquardt}}, \bibinfo {author} {\bibfnamefont {J.~P.}\ \bibnamefont {Chen}},
  \bibinfo {author} {\bibfnamefont {A.~A.}\ \bibnamefont {Clerk}}, \ and\
  \bibinfo {author} {\bibfnamefont {S.~M.}\ \bibnamefont {Girvin}},\ }\href
  {\doibase 10.1103/PhysRevLett.99.093902} {\bibfield  {journal} {\bibinfo
  {journal} {Phys. Rev. Lett.}\ }\textbf {\bibinfo {volume} {99}},\ \bibinfo
  {pages} {093902} (\bibinfo {year} {2007})}\BibitemShut {NoStop}%
\bibitem [{\citenamefont {Teufel}\ \emph {et~al.}(2011)\citenamefont {Teufel},
  \citenamefont {Donner}, \citenamefont {Li}, \citenamefont {Harlow},
  \citenamefont {Allman}, \citenamefont {Cicak}, \citenamefont {Sirois},
  \citenamefont {Whittaker}, \citenamefont {Lehnert},\ and\ \citenamefont
  {Simmonds}}]{teufel2011sideband}%
  \BibitemOpen
  \bibfield  {author} {\bibinfo {author} {\bibfnamefont {J.~D.}\ \bibnamefont
  {Teufel}}, \bibinfo {author} {\bibfnamefont {T.}~\bibnamefont {Donner}},
  \bibinfo {author} {\bibfnamefont {D.}~\bibnamefont {Li}}, \bibinfo {author}
  {\bibfnamefont {J.~W.}\ \bibnamefont {Harlow}}, \bibinfo {author}
  {\bibfnamefont {M.}~\bibnamefont {Allman}}, \bibinfo {author} {\bibfnamefont
  {K.}~\bibnamefont {Cicak}}, \bibinfo {author} {\bibfnamefont {A.~J.}\
  \bibnamefont {Sirois}}, \bibinfo {author} {\bibfnamefont {J.~D.}\
  \bibnamefont {Whittaker}}, \bibinfo {author} {\bibfnamefont {K.~W.}\
  \bibnamefont {Lehnert}}, \ and\ \bibinfo {author} {\bibfnamefont {R.~W.}\
  \bibnamefont {Simmonds}},\ }\href@noop {} {\bibfield  {journal} {\bibinfo
  {journal} {Nature}\ }\textbf {\bibinfo {volume} {475}},\ \bibinfo {pages}
  {359} (\bibinfo {year} {2011})}\BibitemShut {NoStop}%
\bibitem [{\citenamefont {Leibfried}\ \emph {et~al.}(1996)\citenamefont
  {Leibfried}, \citenamefont {Meekhof}, \citenamefont {King}, \citenamefont
  {Monroe}, \citenamefont {Itano},\ and\ \citenamefont
  {Wineland}}]{PhysRevLett.77.4281}%
  \BibitemOpen
  \bibfield  {author} {\bibinfo {author} {\bibfnamefont {D.}~\bibnamefont
  {Leibfried}}, \bibinfo {author} {\bibfnamefont {D.~M.}\ \bibnamefont
  {Meekhof}}, \bibinfo {author} {\bibfnamefont {B.~E.}\ \bibnamefont {King}},
  \bibinfo {author} {\bibfnamefont {C.}~\bibnamefont {Monroe}}, \bibinfo
  {author} {\bibfnamefont {W.~M.}\ \bibnamefont {Itano}}, \ and\ \bibinfo
  {author} {\bibfnamefont {D.~J.}\ \bibnamefont {Wineland}},\ }\href {\doibase
  10.1103/PhysRevLett.77.4281} {\bibfield  {journal} {\bibinfo  {journal}
  {Phys. Rev. Lett.}\ }\textbf {\bibinfo {volume} {77}},\ \bibinfo {pages}
  {4281} (\bibinfo {year} {1996})}\BibitemShut {NoStop}%
\bibitem [{\citenamefont {Lv}\ \emph {et~al.}(2018)\citenamefont {Lv},
  \citenamefont {An}, \citenamefont {Liu}, \citenamefont {Zhang}, \citenamefont
  {Pedernales}, \citenamefont {Lamata}, \citenamefont {Solano},\ and\
  \citenamefont {Kim}}]{PhysRevX.8.021027}%
  \BibitemOpen
  \bibfield  {author} {\bibinfo {author} {\bibfnamefont {D.}~\bibnamefont
  {Lv}}, \bibinfo {author} {\bibfnamefont {S.}~\bibnamefont {An}}, \bibinfo
  {author} {\bibfnamefont {Z.}~\bibnamefont {Liu}}, \bibinfo {author}
  {\bibfnamefont {J.-N.}\ \bibnamefont {Zhang}}, \bibinfo {author}
  {\bibfnamefont {J.~S.}\ \bibnamefont {Pedernales}}, \bibinfo {author}
  {\bibfnamefont {L.}~\bibnamefont {Lamata}}, \bibinfo {author} {\bibfnamefont
  {E.}~\bibnamefont {Solano}}, \ and\ \bibinfo {author} {\bibfnamefont
  {K.}~\bibnamefont {Kim}},\ }\href {\doibase 10.1103/PhysRevX.8.021027}
  {\bibfield  {journal} {\bibinfo  {journal} {Phys. Rev. X}\ }\textbf {\bibinfo
  {volume} {8}},\ \bibinfo {pages} {021027} (\bibinfo {year}
  {2018})}\BibitemShut {NoStop}%
\bibitem [{\citenamefont {Bennett}\ and\ \citenamefont
  {DiVincenzo}(2000)}]{bennett2000quantum}%
  \BibitemOpen
  \bibfield  {author} {\bibinfo {author} {\bibfnamefont {C.~H.}\ \bibnamefont
  {Bennett}}\ and\ \bibinfo {author} {\bibfnamefont {D.~P.}\ \bibnamefont
  {DiVincenzo}},\ }\href@noop {} {\bibfield  {journal} {\bibinfo  {journal}
  {nature}\ }\textbf {\bibinfo {volume} {404}},\ \bibinfo {pages} {247}
  (\bibinfo {year} {2000})}\BibitemShut {NoStop}%
\bibitem [{\citenamefont {Forn-D\'{\i}az}\ \emph {et~al.}(2019)\citenamefont
  {Forn-D\'{\i}az}, \citenamefont {Lamata}, \citenamefont {Rico}, \citenamefont
  {Kono},\ and\ \citenamefont {Solano}}]{RevModPhys.91.025005}%
  \BibitemOpen
  \bibfield  {author} {\bibinfo {author} {\bibfnamefont {P.}~\bibnamefont
  {Forn-D\'{\i}az}}, \bibinfo {author} {\bibfnamefont {L.}~\bibnamefont
  {Lamata}}, \bibinfo {author} {\bibfnamefont {E.}~\bibnamefont {Rico}},
  \bibinfo {author} {\bibfnamefont {J.}~\bibnamefont {Kono}}, \ and\ \bibinfo
  {author} {\bibfnamefont {E.}~\bibnamefont {Solano}},\ }\href@noop {}
  {\bibfield  {journal} {\bibinfo  {journal} {Rev. Mod. Phys.}\ }\textbf
  {\bibinfo {volume} {91}},\ \bibinfo {pages} {025005} (\bibinfo {year}
  {2019})}\BibitemShut {NoStop}%
\bibitem [{\citenamefont {Ge}\ \emph {et~al.}(2019)\citenamefont {Ge},
  \citenamefont {Sawyer}, \citenamefont {Britton}, \citenamefont {Jacobs},
  \citenamefont {Bollinger},\ and\ \citenamefont
  {Foss-Feig}}]{PhysRevLett.122.030501}%
  \BibitemOpen
  \bibfield  {author} {\bibinfo {author} {\bibfnamefont {W.}~\bibnamefont
  {Ge}}, \bibinfo {author} {\bibfnamefont {B.~C.}\ \bibnamefont {Sawyer}},
  \bibinfo {author} {\bibfnamefont {J.~W.}\ \bibnamefont {Britton}}, \bibinfo
  {author} {\bibfnamefont {K.}~\bibnamefont {Jacobs}}, \bibinfo {author}
  {\bibfnamefont {J.~J.}\ \bibnamefont {Bollinger}}, \ and\ \bibinfo {author}
  {\bibfnamefont {M.}~\bibnamefont {Foss-Feig}},\ }\href {\doibase
  10.1103/PhysRevLett.122.030501} {\bibfield  {journal} {\bibinfo  {journal}
  {Phys. Rev. Lett.}\ }\textbf {\bibinfo {volume} {122}},\ \bibinfo {pages}
  {030501} (\bibinfo {year} {2019})}\BibitemShut {NoStop}%
\bibitem [{\citenamefont {Giovannetti}\ \emph {et~al.}(2006)\citenamefont
  {Giovannetti}, \citenamefont {Lloyd},\ and\ \citenamefont
  {Maccone}}]{PhysRevLett.96.010401}%
  \BibitemOpen
  \bibfield  {author} {\bibinfo {author} {\bibfnamefont {V.}~\bibnamefont
  {Giovannetti}}, \bibinfo {author} {\bibfnamefont {S.}~\bibnamefont {Lloyd}},
  \ and\ \bibinfo {author} {\bibfnamefont {L.}~\bibnamefont {Maccone}},\ }\href
  {\doibase 10.1103/PhysRevLett.96.010401} {\bibfield  {journal} {\bibinfo
  {journal} {Phys. Rev. Lett.}\ }\textbf {\bibinfo {volume} {96}},\ \bibinfo
  {pages} {010401} (\bibinfo {year} {2006})}\BibitemShut {NoStop}%
\bibitem [{\citenamefont {Pezz\`e}\ \emph {et~al.}(2018)\citenamefont
  {Pezz\`e}, \citenamefont {Smerzi}, \citenamefont {Oberthaler}, \citenamefont
  {Schmied},\ and\ \citenamefont {Treutlein}}]{RevModPhys.90.035005}%
  \BibitemOpen
  \bibfield  {author} {\bibinfo {author} {\bibfnamefont {L.}~\bibnamefont
  {Pezz\`e}}, \bibinfo {author} {\bibfnamefont {A.}~\bibnamefont {Smerzi}},
  \bibinfo {author} {\bibfnamefont {M.~K.}\ \bibnamefont {Oberthaler}},
  \bibinfo {author} {\bibfnamefont {R.}~\bibnamefont {Schmied}}, \ and\
  \bibinfo {author} {\bibfnamefont {P.}~\bibnamefont {Treutlein}},\ }\href
  {\doibase 10.1103/RevModPhys.90.035005} {\bibfield  {journal} {\bibinfo
  {journal} {Rev. Mod. Phys.}\ }\textbf {\bibinfo {volume} {90}},\ \bibinfo
  {pages} {035005} (\bibinfo {year} {2018})}\BibitemShut {NoStop}%
\bibitem [{\citenamefont {Degen}\ \emph {et~al.}(2017)\citenamefont {Degen},
  \citenamefont {Reinhard},\ and\ \citenamefont
  {Cappellaro}}]{RevModPhys.89.035002}%
  \BibitemOpen
  \bibfield  {author} {\bibinfo {author} {\bibfnamefont {C.~L.}\ \bibnamefont
  {Degen}}, \bibinfo {author} {\bibfnamefont {F.}~\bibnamefont {Reinhard}}, \
  and\ \bibinfo {author} {\bibfnamefont {P.}~\bibnamefont {Cappellaro}},\
  }\href {\doibase 10.1103/RevModPhys.89.035002} {\bibfield  {journal}
  {\bibinfo  {journal} {Rev. Mod. Phys.}\ }\textbf {\bibinfo {volume} {89}},\
  \bibinfo {pages} {035002} (\bibinfo {year} {2017})}\BibitemShut {NoStop}%
\bibitem [{\citenamefont {Wollman}\ \emph {et~al.}(2015)\citenamefont
  {Wollman}, \citenamefont {Lei}, \citenamefont {Weinstein}, \citenamefont
  {Suh}, \citenamefont {Kronwald}, \citenamefont {Marquardt}, \citenamefont
  {Clerk},\ and\ \citenamefont {Schwab}}]{wollman2015quantum}%
  \BibitemOpen
  \bibfield  {author} {\bibinfo {author} {\bibfnamefont {E.~E.}\ \bibnamefont
  {Wollman}}, \bibinfo {author} {\bibfnamefont {C.}~\bibnamefont {Lei}},
  \bibinfo {author} {\bibfnamefont {A.}~\bibnamefont {Weinstein}}, \bibinfo
  {author} {\bibfnamefont {J.}~\bibnamefont {Suh}}, \bibinfo {author}
  {\bibfnamefont {A.}~\bibnamefont {Kronwald}}, \bibinfo {author}
  {\bibfnamefont {F.}~\bibnamefont {Marquardt}}, \bibinfo {author}
  {\bibfnamefont {A.~A.}\ \bibnamefont {Clerk}}, \ and\ \bibinfo {author}
  {\bibfnamefont {K.}~\bibnamefont {Schwab}},\ }\href@noop {} {\bibfield
  {journal} {\bibinfo  {journal} {Science}\ }\textbf {\bibinfo {volume}
  {349}},\ \bibinfo {pages} {952} (\bibinfo {year} {2015})}\BibitemShut
  {NoStop}%
\bibitem [{\citenamefont {Palomaki}\ \emph {et~al.}(2013)\citenamefont
  {Palomaki}, \citenamefont {Teufel}, \citenamefont {Simmonds},\ and\
  \citenamefont {Lehnert}}]{palomaki2013entangling}%
  \BibitemOpen
  \bibfield  {author} {\bibinfo {author} {\bibfnamefont {T.}~\bibnamefont
  {Palomaki}}, \bibinfo {author} {\bibfnamefont {J.}~\bibnamefont {Teufel}},
  \bibinfo {author} {\bibfnamefont {R.}~\bibnamefont {Simmonds}}, \ and\
  \bibinfo {author} {\bibfnamefont {K.~W.}\ \bibnamefont {Lehnert}},\
  }\href@noop {} {\bibfield  {journal} {\bibinfo  {journal} {Science}\ }\textbf
  {\bibinfo {volume} {342}},\ \bibinfo {pages} {710} (\bibinfo {year}
  {2013})}\BibitemShut {NoStop}%
\bibitem [{\citenamefont {Meekhof}\ \emph {et~al.}(1996)\citenamefont
  {Meekhof}, \citenamefont {Monroe}, \citenamefont {King}, \citenamefont
  {Itano},\ and\ \citenamefont {Wineland}}]{PhysRevLett.76.1796}%
  \BibitemOpen
  \bibfield  {author} {\bibinfo {author} {\bibfnamefont {D.~M.}\ \bibnamefont
  {Meekhof}}, \bibinfo {author} {\bibfnamefont {C.}~\bibnamefont {Monroe}},
  \bibinfo {author} {\bibfnamefont {B.~E.}\ \bibnamefont {King}}, \bibinfo
  {author} {\bibfnamefont {W.~M.}\ \bibnamefont {Itano}}, \ and\ \bibinfo
  {author} {\bibfnamefont {D.~J.}\ \bibnamefont {Wineland}},\ }\href {\doibase
  10.1103/PhysRevLett.76.1796} {\bibfield  {journal} {\bibinfo  {journal}
  {Phys. Rev. Lett.}\ }\textbf {\bibinfo {volume} {76}},\ \bibinfo {pages}
  {1796} (\bibinfo {year} {1996})}\BibitemShut {NoStop}%
\bibitem [{\citenamefont {Mu{\~n}oz}\ \emph {et~al.}(2014)\citenamefont
  {Mu{\~n}oz}, \citenamefont {Del~Valle}, \citenamefont {Tudela}, \citenamefont
  {M{\"u}ller}, \citenamefont {Lichtmannecker}, \citenamefont {Kaniber},
  \citenamefont {Tejedor}, \citenamefont {Finley},\ and\ \citenamefont
  {Laussy}}]{munoz2014emitters}%
  \BibitemOpen
  \bibfield  {author} {\bibinfo {author} {\bibfnamefont {C.~S.}\ \bibnamefont
  {Mu{\~n}oz}}, \bibinfo {author} {\bibfnamefont {E.}~\bibnamefont
  {Del~Valle}}, \bibinfo {author} {\bibfnamefont {A.~G.}\ \bibnamefont
  {Tudela}}, \bibinfo {author} {\bibfnamefont {K.}~\bibnamefont {M{\"u}ller}},
  \bibinfo {author} {\bibfnamefont {S.}~\bibnamefont {Lichtmannecker}},
  \bibinfo {author} {\bibfnamefont {M.}~\bibnamefont {Kaniber}}, \bibinfo
  {author} {\bibfnamefont {C.}~\bibnamefont {Tejedor}}, \bibinfo {author}
  {\bibfnamefont {J.}~\bibnamefont {Finley}}, \ and\ \bibinfo {author}
  {\bibfnamefont {F.}~\bibnamefont {Laussy}},\ }\href@noop {} {\bibfield
  {journal} {\bibinfo  {journal} {Nature photonics}\ }\textbf {\bibinfo
  {volume} {8}},\ \bibinfo {pages} {550} (\bibinfo {year} {2014})}\BibitemShut
  {NoStop}%
\bibitem [{\citenamefont {Chang}\ \emph {et~al.}(2016)\citenamefont {Chang},
  \citenamefont {Gonz\'alez-Tudela}, \citenamefont {S\'anchez Mu\~noz},
  \citenamefont {Navarrete-Benlloch},\ and\ \citenamefont
  {Shi}}]{PhysRevLett.117.203602}%
  \BibitemOpen
  \bibfield  {author} {\bibinfo {author} {\bibfnamefont {Y.}~\bibnamefont
  {Chang}}, \bibinfo {author} {\bibfnamefont {A.}~\bibnamefont
  {Gonz\'alez-Tudela}}, \bibinfo {author} {\bibfnamefont {C.}~\bibnamefont
  {S\'anchez Mu\~noz}}, \bibinfo {author} {\bibfnamefont {C.}~\bibnamefont
  {Navarrete-Benlloch}}, \ and\ \bibinfo {author} {\bibfnamefont
  {T.}~\bibnamefont {Shi}},\ }\href {\doibase 10.1103/PhysRevLett.117.203602}
  {\bibfield  {journal} {\bibinfo  {journal} {Phys. Rev. Lett.}\ }\textbf
  {\bibinfo {volume} {117}},\ \bibinfo {pages} {203602} (\bibinfo {year}
  {2016})}\BibitemShut {NoStop}%
\bibitem [{\citenamefont {Bin}\ \emph {et~al.}(2020)\citenamefont {Bin},
  \citenamefont {L\"u}, \citenamefont {Laussy}, \citenamefont {Nori},\ and\
  \citenamefont {Wu}}]{PhysRevLett.124.053601}%
  \BibitemOpen
  \bibfield  {author} {\bibinfo {author} {\bibfnamefont {Q.}~\bibnamefont
  {Bin}}, \bibinfo {author} {\bibfnamefont {X.-Y.}\ \bibnamefont {L\"u}},
  \bibinfo {author} {\bibfnamefont {F.~P.}\ \bibnamefont {Laussy}}, \bibinfo
  {author} {\bibfnamefont {F.}~\bibnamefont {Nori}}, \ and\ \bibinfo {author}
  {\bibfnamefont {Y.}~\bibnamefont {Wu}},\ }\href {\doibase
  10.1103/PhysRevLett.124.053601} {\bibfield  {journal} {\bibinfo  {journal}
  {Phys. Rev. Lett.}\ }\textbf {\bibinfo {volume} {124}},\ \bibinfo {pages}
  {053601} (\bibinfo {year} {2020})}\BibitemShut {NoStop}%
\bibitem [{\citenamefont {Bin}\ \emph {et~al.}(2021)\citenamefont {Bin},
  \citenamefont {Wu},\ and\ \citenamefont {L\"u}}]{PhysRevLett.127.073602}%
  \BibitemOpen
  \bibfield  {author} {\bibinfo {author} {\bibfnamefont {Q.}~\bibnamefont
  {Bin}}, \bibinfo {author} {\bibfnamefont {Y.}~\bibnamefont {Wu}}, \ and\
  \bibinfo {author} {\bibfnamefont {X.-Y.}\ \bibnamefont {L\"u}},\ }\href
  {\doibase 10.1103/PhysRevLett.127.073602} {\bibfield  {journal} {\bibinfo
  {journal} {Phys. Rev. Lett.}\ }\textbf {\bibinfo {volume} {127}},\ \bibinfo
  {pages} {073602} (\bibinfo {year} {2021})}\BibitemShut {NoStop}%
\bibitem [{\citenamefont {Deng}\ \emph {et~al.}(2021)\citenamefont {Deng},
  \citenamefont {Shi},\ and\ \citenamefont {Yi}}]{deng2021motional}%
  \BibitemOpen
  \bibfield  {author} {\bibinfo {author} {\bibfnamefont {Y.}~\bibnamefont
  {Deng}}, \bibinfo {author} {\bibfnamefont {T.}~\bibnamefont {Shi}}, \ and\
  \bibinfo {author} {\bibfnamefont {S.}~\bibnamefont {Yi}},\ }\href@noop {}
  {\bibfield  {journal} {\bibinfo  {journal} {Photonics Research}\ }\textbf
  {\bibinfo {volume} {9}},\ \bibinfo {pages} {1289} (\bibinfo {year}
  {2021})}\BibitemShut {NoStop}%
\bibitem [{\citenamefont {Mirhosseini}\ \emph {et~al.}(2020)\citenamefont
  {Mirhosseini}, \citenamefont {Sipahigil}, \citenamefont {Kalaee},\ and\
  \citenamefont {Painter}}]{mirhosseini2020superconducting}%
  \BibitemOpen
  \bibfield  {author} {\bibinfo {author} {\bibfnamefont {M.}~\bibnamefont
  {Mirhosseini}}, \bibinfo {author} {\bibfnamefont {A.}~\bibnamefont
  {Sipahigil}}, \bibinfo {author} {\bibfnamefont {M.}~\bibnamefont {Kalaee}}, \
  and\ \bibinfo {author} {\bibfnamefont {O.}~\bibnamefont {Painter}},\
  }\href@noop {} {\bibfield  {journal} {\bibinfo  {journal} {Nature}\ }\textbf
  {\bibinfo {volume} {588}},\ \bibinfo {pages} {599} (\bibinfo {year}
  {2020})}\BibitemShut {NoStop}%
\bibitem [{\citenamefont {Briegel}\ \emph {et~al.}(1998)\citenamefont
  {Briegel}, \citenamefont {D\"ur}, \citenamefont {Cirac},\ and\ \citenamefont
  {Zoller}}]{PhysRevLett.81.5932}%
  \BibitemOpen
  \bibfield  {author} {\bibinfo {author} {\bibfnamefont {H.-J.}\ \bibnamefont
  {Briegel}}, \bibinfo {author} {\bibfnamefont {W.}~\bibnamefont {D\"ur}},
  \bibinfo {author} {\bibfnamefont {J.~I.}\ \bibnamefont {Cirac}}, \ and\
  \bibinfo {author} {\bibfnamefont {P.}~\bibnamefont {Zoller}},\ }\href
  {\doibase 10.1103/PhysRevLett.81.5932} {\bibfield  {journal} {\bibinfo
  {journal} {Phys. Rev. Lett.}\ }\textbf {\bibinfo {volume} {81}},\ \bibinfo
  {pages} {5932} (\bibinfo {year} {1998})}\BibitemShut {NoStop}%
\bibitem [{\citenamefont {Forsch}\ \emph {et~al.}(2020)\citenamefont {Forsch},
  \citenamefont {Stockill}, \citenamefont {Wallucks}, \citenamefont
  {Marinkovi{\'c}}, \citenamefont {G{\"a}rtner}, \citenamefont {Norte},
  \citenamefont {van Otten}, \citenamefont {Fiore}, \citenamefont
  {Srinivasan},\ and\ \citenamefont {Gr{\"o}blacher}}]{forsch2020microwave}%
  \BibitemOpen
  \bibfield  {author} {\bibinfo {author} {\bibfnamefont {M.}~\bibnamefont
  {Forsch}}, \bibinfo {author} {\bibfnamefont {R.}~\bibnamefont {Stockill}},
  \bibinfo {author} {\bibfnamefont {A.}~\bibnamefont {Wallucks}}, \bibinfo
  {author} {\bibfnamefont {I.}~\bibnamefont {Marinkovi{\'c}}}, \bibinfo
  {author} {\bibfnamefont {C.}~\bibnamefont {G{\"a}rtner}}, \bibinfo {author}
  {\bibfnamefont {R.~A.}\ \bibnamefont {Norte}}, \bibinfo {author}
  {\bibfnamefont {F.}~\bibnamefont {van Otten}}, \bibinfo {author}
  {\bibfnamefont {A.}~\bibnamefont {Fiore}}, \bibinfo {author} {\bibfnamefont
  {K.}~\bibnamefont {Srinivasan}}, \ and\ \bibinfo {author} {\bibfnamefont
  {S.}~\bibnamefont {Gr{\"o}blacher}},\ }\href@noop {} {\bibfield  {journal}
  {\bibinfo  {journal} {Nature Physics}\ }\textbf {\bibinfo {volume} {16}},\
  \bibinfo {pages} {69} (\bibinfo {year} {2020})}\BibitemShut {NoStop}%
\bibitem [{\citenamefont {Bagci}\ \emph {et~al.}(2014)\citenamefont {Bagci},
  \citenamefont {Simonsen}, \citenamefont {Schmid}, \citenamefont {Villanueva},
  \citenamefont {Zeuthen}, \citenamefont {Appel}, \citenamefont {Taylor},
  \citenamefont {S{\o}rensen}, \citenamefont {Usami}, \citenamefont
  {Schliesser} \emph {et~al.}}]{bagci2014optical}%
  \BibitemOpen
  \bibfield  {author} {\bibinfo {author} {\bibfnamefont {T.}~\bibnamefont
  {Bagci}}, \bibinfo {author} {\bibfnamefont {A.}~\bibnamefont {Simonsen}},
  \bibinfo {author} {\bibfnamefont {S.}~\bibnamefont {Schmid}}, \bibinfo
  {author} {\bibfnamefont {L.~G.}\ \bibnamefont {Villanueva}}, \bibinfo
  {author} {\bibfnamefont {E.}~\bibnamefont {Zeuthen}}, \bibinfo {author}
  {\bibfnamefont {J.}~\bibnamefont {Appel}}, \bibinfo {author} {\bibfnamefont
  {J.~M.}\ \bibnamefont {Taylor}}, \bibinfo {author} {\bibfnamefont
  {A.}~\bibnamefont {S{\o}rensen}}, \bibinfo {author} {\bibfnamefont
  {K.}~\bibnamefont {Usami}}, \bibinfo {author} {\bibfnamefont
  {A.}~\bibnamefont {Schliesser}},  \emph {et~al.},\ }\href@noop {} {\bibfield
  {journal} {\bibinfo  {journal} {Nature}\ }\textbf {\bibinfo {volume} {507}},\
  \bibinfo {pages} {81} (\bibinfo {year} {2014})}\BibitemShut {NoStop}%
\bibitem [{\citenamefont {Bochmann}\ \emph {et~al.}(2013)\citenamefont
  {Bochmann}, \citenamefont {Vainsencher}, \citenamefont {Awschalom},\ and\
  \citenamefont {Cleland}}]{bochmann2013nanomechanical}%
  \BibitemOpen
  \bibfield  {author} {\bibinfo {author} {\bibfnamefont {J.}~\bibnamefont
  {Bochmann}}, \bibinfo {author} {\bibfnamefont {A.}~\bibnamefont
  {Vainsencher}}, \bibinfo {author} {\bibfnamefont {D.~D.}\ \bibnamefont
  {Awschalom}}, \ and\ \bibinfo {author} {\bibfnamefont {A.~N.}\ \bibnamefont
  {Cleland}},\ }\href@noop {} {\bibfield  {journal} {\bibinfo  {journal}
  {Nature Physics}\ }\textbf {\bibinfo {volume} {9}},\ \bibinfo {pages} {712}
  (\bibinfo {year} {2013})}\BibitemShut {NoStop}%
\bibitem [{\citenamefont {Schuetz}\ \emph {et~al.}(2015)\citenamefont
  {Schuetz}, \citenamefont {Kessler}, \citenamefont {Giedke}, \citenamefont
  {Vandersypen}, \citenamefont {Lukin},\ and\ \citenamefont
  {Cirac}}]{PhysRevX.5.031031}%
  \BibitemOpen
  \bibfield  {author} {\bibinfo {author} {\bibfnamefont {M.~J.~A.}\
  \bibnamefont {Schuetz}}, \bibinfo {author} {\bibfnamefont {E.~M.}\
  \bibnamefont {Kessler}}, \bibinfo {author} {\bibfnamefont {G.}~\bibnamefont
  {Giedke}}, \bibinfo {author} {\bibfnamefont {L.~M.~K.}\ \bibnamefont
  {Vandersypen}}, \bibinfo {author} {\bibfnamefont {M.~D.}\ \bibnamefont
  {Lukin}}, \ and\ \bibinfo {author} {\bibfnamefont {J.~I.}\ \bibnamefont
  {Cirac}},\ }\href {\doibase 10.1103/PhysRevX.5.031031} {\bibfield  {journal}
  {\bibinfo  {journal} {Phys. Rev. X}\ }\textbf {\bibinfo {volume} {5}},\
  \bibinfo {pages} {031031} (\bibinfo {year} {2015})}\BibitemShut {NoStop}%
\bibitem [{\citenamefont {Noguchi}\ \emph {et~al.}(2017)\citenamefont
  {Noguchi}, \citenamefont {Yamazaki}, \citenamefont {Tabuchi},\ and\
  \citenamefont {Nakamura}}]{PhysRevLett.119.180505}%
  \BibitemOpen
  \bibfield  {author} {\bibinfo {author} {\bibfnamefont {A.}~\bibnamefont
  {Noguchi}}, \bibinfo {author} {\bibfnamefont {R.}~\bibnamefont {Yamazaki}},
  \bibinfo {author} {\bibfnamefont {Y.}~\bibnamefont {Tabuchi}}, \ and\
  \bibinfo {author} {\bibfnamefont {Y.}~\bibnamefont {Nakamura}},\ }\href
  {\doibase 10.1103/PhysRevLett.119.180505} {\bibfield  {journal} {\bibinfo
  {journal} {Phys. Rev. Lett.}\ }\textbf {\bibinfo {volume} {119}},\ \bibinfo
  {pages} {180505} (\bibinfo {year} {2017})}\BibitemShut {NoStop}%
\bibitem [{\citenamefont {Mari}\ and\ \citenamefont
  {Eisert}(2012)}]{PhysRevLett.109.230503}%
  \BibitemOpen
  \bibfield  {author} {\bibinfo {author} {\bibfnamefont {A.}~\bibnamefont
  {Mari}}\ and\ \bibinfo {author} {\bibfnamefont {J.}~\bibnamefont {Eisert}},\
  }\href {\doibase 10.1103/PhysRevLett.109.230503} {\bibfield  {journal}
  {\bibinfo  {journal} {Phys. Rev. Lett.}\ }\textbf {\bibinfo {volume} {109}},\
  \bibinfo {pages} {230503} (\bibinfo {year} {2012})}\BibitemShut {NoStop}%
\bibitem [{\citenamefont {Arvidsson-Shukur}\ \emph {et~al.}(2020)\citenamefont
  {Arvidsson-Shukur}, \citenamefont {Yunger~Halpern}, \citenamefont {Lepage},
  \citenamefont {Lasek}, \citenamefont {Barnes},\ and\ \citenamefont
  {Lloyd}}]{arvidsson2020quantum}%
  \BibitemOpen
  \bibfield  {author} {\bibinfo {author} {\bibfnamefont {D.~R.}\ \bibnamefont
  {Arvidsson-Shukur}}, \bibinfo {author} {\bibfnamefont {N.}~\bibnamefont
  {Yunger~Halpern}}, \bibinfo {author} {\bibfnamefont {H.~V.}\ \bibnamefont
  {Lepage}}, \bibinfo {author} {\bibfnamefont {A.~A.}\ \bibnamefont {Lasek}},
  \bibinfo {author} {\bibfnamefont {C.~H.}\ \bibnamefont {Barnes}}, \ and\
  \bibinfo {author} {\bibfnamefont {S.}~\bibnamefont {Lloyd}},\ }\href@noop {}
  {\bibfield  {journal} {\bibinfo  {journal} {Nature communications}\ }\textbf
  {\bibinfo {volume} {11}},\ \bibinfo {pages} {3775} (\bibinfo {year}
  {2020})}\BibitemShut {NoStop}%
\bibitem [{\citenamefont {Marinkovi\ifmmode~\acute{c}\else \'{c}\fi{}}\ \emph
  {et~al.}(2018)\citenamefont {Marinkovi\ifmmode~\acute{c}\else \'{c}\fi{}},
  \citenamefont {Wallucks}, \citenamefont {Riedinger}, \citenamefont {Hong},
  \citenamefont {Aspelmeyer},\ and\ \citenamefont
  {Gr\"oblacher}}]{PhysRevLett.121.220404}%
  \BibitemOpen
  \bibfield  {author} {\bibinfo {author} {\bibfnamefont {I.}~\bibnamefont
  {Marinkovi\ifmmode~\acute{c}\else \'{c}\fi{}}}, \bibinfo {author}
  {\bibfnamefont {A.}~\bibnamefont {Wallucks}}, \bibinfo {author}
  {\bibfnamefont {R.}~\bibnamefont {Riedinger}}, \bibinfo {author}
  {\bibfnamefont {S.}~\bibnamefont {Hong}}, \bibinfo {author} {\bibfnamefont
  {M.}~\bibnamefont {Aspelmeyer}}, \ and\ \bibinfo {author} {\bibfnamefont
  {S.}~\bibnamefont {Gr\"oblacher}},\ }\href {\doibase
  10.1103/PhysRevLett.121.220404} {\bibfield  {journal} {\bibinfo  {journal}
  {Phys. Rev. Lett.}\ }\textbf {\bibinfo {volume} {121}},\ \bibinfo {pages}
  {220404} (\bibinfo {year} {2018})}\BibitemShut {NoStop}%
\bibitem [{\citenamefont {Miao}\ \emph {et~al.}(2015)\citenamefont {Miao},
  \citenamefont {Ma}, \citenamefont {Zhao},\ and\ \citenamefont
  {Chen}}]{PhysRevLett.115.211104}%
  \BibitemOpen
  \bibfield  {author} {\bibinfo {author} {\bibfnamefont {H.}~\bibnamefont
  {Miao}}, \bibinfo {author} {\bibfnamefont {Y.}~\bibnamefont {Ma}}, \bibinfo
  {author} {\bibfnamefont {C.}~\bibnamefont {Zhao}}, \ and\ \bibinfo {author}
  {\bibfnamefont {Y.}~\bibnamefont {Chen}},\ }\href {\doibase
  10.1103/PhysRevLett.115.211104} {\bibfield  {journal} {\bibinfo  {journal}
  {Phys. Rev. Lett.}\ }\textbf {\bibinfo {volume} {115}},\ \bibinfo {pages}
  {211104} (\bibinfo {year} {2015})}\BibitemShut {NoStop}%
\bibitem [{\citenamefont {Korobko}\ \emph {et~al.}(2017)\citenamefont
  {Korobko}, \citenamefont {Kleybolte}, \citenamefont {Ast}, \citenamefont
  {Miao}, \citenamefont {Chen},\ and\ \citenamefont
  {Schnabel}}]{PhysRevLett.118.143601}%
  \BibitemOpen
  \bibfield  {author} {\bibinfo {author} {\bibfnamefont {M.}~\bibnamefont
  {Korobko}}, \bibinfo {author} {\bibfnamefont {L.}~\bibnamefont {Kleybolte}},
  \bibinfo {author} {\bibfnamefont {S.}~\bibnamefont {Ast}}, \bibinfo {author}
  {\bibfnamefont {H.}~\bibnamefont {Miao}}, \bibinfo {author} {\bibfnamefont
  {Y.}~\bibnamefont {Chen}}, \ and\ \bibinfo {author} {\bibfnamefont
  {R.}~\bibnamefont {Schnabel}},\ }\href {\doibase
  10.1103/PhysRevLett.118.143601} {\bibfield  {journal} {\bibinfo  {journal}
  {Phys. Rev. Lett.}\ }\textbf {\bibinfo {volume} {118}},\ \bibinfo {pages}
  {143601} (\bibinfo {year} {2017})}\BibitemShut {NoStop}%
\bibitem [{\citenamefont {Cata\~no Lopez}\ \emph {et~al.}(2020)\citenamefont
  {Cata\~no Lopez}, \citenamefont {Santiago-Condori}, \citenamefont
  {Edamatsu},\ and\ \citenamefont {Matsumoto}}]{PhysRevLett.124.221102}%
  \BibitemOpen
  \bibfield  {author} {\bibinfo {author} {\bibfnamefont {S.~B.}\ \bibnamefont
  {Cata\~no Lopez}}, \bibinfo {author} {\bibfnamefont {J.~G.}\ \bibnamefont
  {Santiago-Condori}}, \bibinfo {author} {\bibfnamefont {K.}~\bibnamefont
  {Edamatsu}}, \ and\ \bibinfo {author} {\bibfnamefont {N.}~\bibnamefont
  {Matsumoto}},\ }\href {\doibase 10.1103/PhysRevLett.124.221102} {\bibfield
  {journal} {\bibinfo  {journal} {Phys. Rev. Lett.}\ }\textbf {\bibinfo
  {volume} {124}},\ \bibinfo {pages} {221102} (\bibinfo {year}
  {2020})}\BibitemShut {NoStop}%
\bibitem [{\citenamefont {Kimble}(2008)}]{kimble2008quantum}%
  \BibitemOpen
  \bibfield  {author} {\bibinfo {author} {\bibfnamefont {H.~J.}\ \bibnamefont
  {Kimble}},\ }\href@noop {} {\bibfield  {journal} {\bibinfo  {journal}
  {Nature}\ }\textbf {\bibinfo {volume} {453}},\ \bibinfo {pages} {1023}
  (\bibinfo {year} {2008})}\BibitemShut {NoStop}%
\bibitem [{\citenamefont {Rabi}(1936)}]{PhysRev.49.324}%
  \BibitemOpen
  \bibfield  {author} {\bibinfo {author} {\bibfnamefont {I.~I.}\ \bibnamefont
  {Rabi}},\ }\href {\doibase 10.1103/PhysRev.49.324} {\bibfield  {journal}
  {\bibinfo  {journal} {Phys. Rev.}\ }\textbf {\bibinfo {volume} {49}},\
  \bibinfo {pages} {324} (\bibinfo {year} {1936})}\BibitemShut {NoStop}%
\bibitem [{\citenamefont {Jaynes}\ and\ \citenamefont
  {Cummings}(1963)}]{jaynes1963comparison}%
  \BibitemOpen
  \bibfield  {author} {\bibinfo {author} {\bibfnamefont {E.~T.}\ \bibnamefont
  {Jaynes}}\ and\ \bibinfo {author} {\bibfnamefont {F.~W.}\ \bibnamefont
  {Cummings}},\ }\href@noop {} {\bibfield  {journal} {\bibinfo  {journal}
  {Proceedings of the IEEE}\ }\textbf {\bibinfo {volume} {51}},\ \bibinfo
  {pages} {89} (\bibinfo {year} {1963})}\BibitemShut {NoStop}%
\bibitem [{\citenamefont {O?Connell}\ \emph {et~al.}(2010)\citenamefont
  {O?Connell}, \citenamefont {Hofheinz}, \citenamefont {Ansmann}, \citenamefont
  {Bialczak}, \citenamefont {Lenander}, \citenamefont {Lucero}, \citenamefont
  {Neeley}, \citenamefont {Sank}, \citenamefont {Wang}, \citenamefont {Weides}
  \emph {et~al.}}]{o2010quantum}%
  \BibitemOpen
  \bibfield  {author} {\bibinfo {author} {\bibfnamefont {A.~D.}\ \bibnamefont
  {O?Connell}}, \bibinfo {author} {\bibfnamefont {M.}~\bibnamefont {Hofheinz}},
  \bibinfo {author} {\bibfnamefont {M.}~\bibnamefont {Ansmann}}, \bibinfo
  {author} {\bibfnamefont {R.~C.}\ \bibnamefont {Bialczak}}, \bibinfo {author}
  {\bibfnamefont {M.}~\bibnamefont {Lenander}}, \bibinfo {author}
  {\bibfnamefont {E.}~\bibnamefont {Lucero}}, \bibinfo {author} {\bibfnamefont
  {M.}~\bibnamefont {Neeley}}, \bibinfo {author} {\bibfnamefont
  {D.}~\bibnamefont {Sank}}, \bibinfo {author} {\bibfnamefont {H.}~\bibnamefont
  {Wang}}, \bibinfo {author} {\bibfnamefont {M.}~\bibnamefont {Weides}},  \emph
  {et~al.},\ }\href@noop {} {\bibfield  {journal} {\bibinfo  {journal}
  {Nature}\ }\textbf {\bibinfo {volume} {464}},\ \bibinfo {pages} {697}
  (\bibinfo {year} {2010})}\BibitemShut {NoStop}%
\bibitem [{\citenamefont {Chu}\ \emph {et~al.}(2018)\citenamefont {Chu},
  \citenamefont {Kharel}, \citenamefont {Yoon}, \citenamefont {Frunzio},
  \citenamefont {Rakich},\ and\ \citenamefont {Schoelkopf}}]{chu2018creation}%
  \BibitemOpen
  \bibfield  {author} {\bibinfo {author} {\bibfnamefont {Y.}~\bibnamefont
  {Chu}}, \bibinfo {author} {\bibfnamefont {P.}~\bibnamefont {Kharel}},
  \bibinfo {author} {\bibfnamefont {T.}~\bibnamefont {Yoon}}, \bibinfo {author}
  {\bibfnamefont {L.}~\bibnamefont {Frunzio}}, \bibinfo {author} {\bibfnamefont
  {P.~T.}\ \bibnamefont {Rakich}}, \ and\ \bibinfo {author} {\bibfnamefont
  {R.~J.}\ \bibnamefont {Schoelkopf}},\ }\href@noop {} {\bibfield  {journal}
  {\bibinfo  {journal} {Nature}\ }\textbf {\bibinfo {volume} {563}},\ \bibinfo
  {pages} {666} (\bibinfo {year} {2018})}\BibitemShut {NoStop}%
\bibitem [{\citenamefont {Arrangoiz-Arriola}\ \emph {et~al.}(2019)\citenamefont
  {Arrangoiz-Arriola}, \citenamefont {Wollack}, \citenamefont {Wang},
  \citenamefont {Pechal}, \citenamefont {Jiang}, \citenamefont {McKenna},
  \citenamefont {Witmer}, \citenamefont {Van~Laer},\ and\ \citenamefont
  {Safavi-Naeini}}]{arrangoiz2019resolving}%
  \BibitemOpen
  \bibfield  {author} {\bibinfo {author} {\bibfnamefont {P.}~\bibnamefont
  {Arrangoiz-Arriola}}, \bibinfo {author} {\bibfnamefont {E.~A.}\ \bibnamefont
  {Wollack}}, \bibinfo {author} {\bibfnamefont {Z.}~\bibnamefont {Wang}},
  \bibinfo {author} {\bibfnamefont {M.}~\bibnamefont {Pechal}}, \bibinfo
  {author} {\bibfnamefont {W.}~\bibnamefont {Jiang}}, \bibinfo {author}
  {\bibfnamefont {T.~P.}\ \bibnamefont {McKenna}}, \bibinfo {author}
  {\bibfnamefont {J.~D.}\ \bibnamefont {Witmer}}, \bibinfo {author}
  {\bibfnamefont {R.}~\bibnamefont {Van~Laer}}, \ and\ \bibinfo {author}
  {\bibfnamefont {A.~H.}\ \bibnamefont {Safavi-Naeini}},\ }\href@noop {}
  {\bibfield  {journal} {\bibinfo  {journal} {Nature}\ }\textbf {\bibinfo
  {volume} {571}},\ \bibinfo {pages} {537} (\bibinfo {year}
  {2019})}\BibitemShut {NoStop}%
\bibitem [{\citenamefont {Sletten}\ \emph {et~al.}(2019)\citenamefont
  {Sletten}, \citenamefont {Moores}, \citenamefont {Viennot},\ and\
  \citenamefont {Lehnert}}]{PhysRevX.9.021056}%
  \BibitemOpen
  \bibfield  {author} {\bibinfo {author} {\bibfnamefont {L.~R.}\ \bibnamefont
  {Sletten}}, \bibinfo {author} {\bibfnamefont {B.~A.}\ \bibnamefont {Moores}},
  \bibinfo {author} {\bibfnamefont {J.~J.}\ \bibnamefont {Viennot}}, \ and\
  \bibinfo {author} {\bibfnamefont {K.~W.}\ \bibnamefont {Lehnert}},\ }\href
  {\doibase 10.1103/PhysRevX.9.021056} {\bibfield  {journal} {\bibinfo
  {journal} {Phys. Rev. X}\ }\textbf {\bibinfo {volume} {9}},\ \bibinfo {pages}
  {021056} (\bibinfo {year} {2019})}\BibitemShut {NoStop}%
\bibitem [{\citenamefont {Patil}\ \emph {et~al.}(2022)\citenamefont {Patil},
  \citenamefont {Yu}, \citenamefont {Frazier}, \citenamefont {Wang},
  \citenamefont {Johnson}, \citenamefont {Fox}, \citenamefont {Reichel},\ and\
  \citenamefont {Harris}}]{PhysRevLett.128.183601}%
  \BibitemOpen
  \bibfield  {author} {\bibinfo {author} {\bibfnamefont {Y.~S.~S.}\
  \bibnamefont {Patil}}, \bibinfo {author} {\bibfnamefont {J.}~\bibnamefont
  {Yu}}, \bibinfo {author} {\bibfnamefont {S.}~\bibnamefont {Frazier}},
  \bibinfo {author} {\bibfnamefont {Y.}~\bibnamefont {Wang}}, \bibinfo {author}
  {\bibfnamefont {K.}~\bibnamefont {Johnson}}, \bibinfo {author} {\bibfnamefont
  {J.}~\bibnamefont {Fox}}, \bibinfo {author} {\bibfnamefont {J.}~\bibnamefont
  {Reichel}}, \ and\ \bibinfo {author} {\bibfnamefont {J.~G.~E.}\ \bibnamefont
  {Harris}},\ }\href {\doibase 10.1103/PhysRevLett.128.183601} {\bibfield
  {journal} {\bibinfo  {journal} {Phys. Rev. Lett.}\ }\textbf {\bibinfo
  {volume} {128}},\ \bibinfo {pages} {183601} (\bibinfo {year}
  {2022})}\BibitemShut {NoStop}%
\bibitem [{\citenamefont {Riedinger}\ \emph {et~al.}(2016)\citenamefont
  {Riedinger}, \citenamefont {Hong}, \citenamefont {Norte}, \citenamefont
  {Slater}, \citenamefont {Shang}, \citenamefont {Krause}, \citenamefont
  {Anant}, \citenamefont {Aspelmeyer},\ and\ \citenamefont
  {Gr{\"o}blacher}}]{riedinger2016non}%
  \BibitemOpen
  \bibfield  {author} {\bibinfo {author} {\bibfnamefont {R.}~\bibnamefont
  {Riedinger}}, \bibinfo {author} {\bibfnamefont {S.}~\bibnamefont {Hong}},
  \bibinfo {author} {\bibfnamefont {R.~A.}\ \bibnamefont {Norte}}, \bibinfo
  {author} {\bibfnamefont {J.~A.}\ \bibnamefont {Slater}}, \bibinfo {author}
  {\bibfnamefont {J.}~\bibnamefont {Shang}}, \bibinfo {author} {\bibfnamefont
  {A.~G.}\ \bibnamefont {Krause}}, \bibinfo {author} {\bibfnamefont
  {V.}~\bibnamefont {Anant}}, \bibinfo {author} {\bibfnamefont
  {M.}~\bibnamefont {Aspelmeyer}}, \ and\ \bibinfo {author} {\bibfnamefont
  {S.}~\bibnamefont {Gr{\"o}blacher}},\ }\href@noop {} {\bibfield  {journal}
  {\bibinfo  {journal} {Nature}\ }\textbf {\bibinfo {volume} {530}},\ \bibinfo
  {pages} {313} (\bibinfo {year} {2016})}\BibitemShut {NoStop}%
\bibitem [{\citenamefont {Hong}\ \emph {et~al.}(2017)\citenamefont {Hong},
  \citenamefont {Riedinger}, \citenamefont {Marinkovi{\'c}}, \citenamefont
  {Wallucks}, \citenamefont {Hofer}, \citenamefont {Norte}, \citenamefont
  {Aspelmeyer},\ and\ \citenamefont {Gr{\"o}blacher}}]{hong2017hanbury}%
  \BibitemOpen
  \bibfield  {author} {\bibinfo {author} {\bibfnamefont {S.}~\bibnamefont
  {Hong}}, \bibinfo {author} {\bibfnamefont {R.}~\bibnamefont {Riedinger}},
  \bibinfo {author} {\bibfnamefont {I.}~\bibnamefont {Marinkovi{\'c}}},
  \bibinfo {author} {\bibfnamefont {A.}~\bibnamefont {Wallucks}}, \bibinfo
  {author} {\bibfnamefont {S.~G.}\ \bibnamefont {Hofer}}, \bibinfo {author}
  {\bibfnamefont {R.~A.}\ \bibnamefont {Norte}}, \bibinfo {author}
  {\bibfnamefont {M.}~\bibnamefont {Aspelmeyer}}, \ and\ \bibinfo {author}
  {\bibfnamefont {S.}~\bibnamefont {Gr{\"o}blacher}},\ }\href@noop {}
  {\bibfield  {journal} {\bibinfo  {journal} {Science}\ }\textbf {\bibinfo
  {volume} {358}},\ \bibinfo {pages} {203} (\bibinfo {year}
  {2017})}\BibitemShut {NoStop}%
\bibitem [{\citenamefont {Riedinger}\ \emph {et~al.}(2018)\citenamefont
  {Riedinger}, \citenamefont {Wallucks}, \citenamefont {Marinkovi{\'c}},
  \citenamefont {L{\"o}schnauer}, \citenamefont {Aspelmeyer}, \citenamefont
  {Hong},\ and\ \citenamefont {Gr{\"o}blacher}}]{riedinger2018remote}%
  \BibitemOpen
  \bibfield  {author} {\bibinfo {author} {\bibfnamefont {R.}~\bibnamefont
  {Riedinger}}, \bibinfo {author} {\bibfnamefont {A.}~\bibnamefont {Wallucks}},
  \bibinfo {author} {\bibfnamefont {I.}~\bibnamefont {Marinkovi{\'c}}},
  \bibinfo {author} {\bibfnamefont {C.}~\bibnamefont {L{\"o}schnauer}},
  \bibinfo {author} {\bibfnamefont {M.}~\bibnamefont {Aspelmeyer}}, \bibinfo
  {author} {\bibfnamefont {S.}~\bibnamefont {Hong}}, \ and\ \bibinfo {author}
  {\bibfnamefont {S.}~\bibnamefont {Gr{\"o}blacher}},\ }\href@noop {}
  {\bibfield  {journal} {\bibinfo  {journal} {Nature}\ }\textbf {\bibinfo
  {volume} {556}},\ \bibinfo {pages} {473} (\bibinfo {year}
  {2018})}\BibitemShut {NoStop}%
\bibitem [{\citenamefont {Lance}\ \emph {et~al.}(2004)\citenamefont {Lance},
  \citenamefont {Symul}, \citenamefont {Bowen}, \citenamefont {Sanders},\ and\
  \citenamefont {Lam}}]{PhysRevLett.92.177903}%
  \BibitemOpen
  \bibfield  {author} {\bibinfo {author} {\bibfnamefont {A.~M.}\ \bibnamefont
  {Lance}}, \bibinfo {author} {\bibfnamefont {T.}~\bibnamefont {Symul}},
  \bibinfo {author} {\bibfnamefont {W.~P.}\ \bibnamefont {Bowen}}, \bibinfo
  {author} {\bibfnamefont {B.~C.}\ \bibnamefont {Sanders}}, \ and\ \bibinfo
  {author} {\bibfnamefont {P.~K.}\ \bibnamefont {Lam}},\ }\href {\doibase
  10.1103/PhysRevLett.92.177903} {\bibfield  {journal} {\bibinfo  {journal}
  {Phys. Rev. Lett.}\ }\textbf {\bibinfo {volume} {92}},\ \bibinfo {pages}
  {177903} (\bibinfo {year} {2004})}\BibitemShut {NoStop}%
\bibitem [{\citenamefont {Jing}\ \emph {et~al.}(2003)\citenamefont {Jing},
  \citenamefont {Zhang}, \citenamefont {Yan}, \citenamefont {Zhao},
  \citenamefont {Xie},\ and\ \citenamefont {Peng}}]{PhysRevLett.90.167903}%
  \BibitemOpen
  \bibfield  {author} {\bibinfo {author} {\bibfnamefont {J.}~\bibnamefont
  {Jing}}, \bibinfo {author} {\bibfnamefont {J.}~\bibnamefont {Zhang}},
  \bibinfo {author} {\bibfnamefont {Y.}~\bibnamefont {Yan}}, \bibinfo {author}
  {\bibfnamefont {F.}~\bibnamefont {Zhao}}, \bibinfo {author} {\bibfnamefont
  {C.}~\bibnamefont {Xie}}, \ and\ \bibinfo {author} {\bibfnamefont
  {K.}~\bibnamefont {Peng}},\ }\href {\doibase 10.1103/PhysRevLett.90.167903}
  {\bibfield  {journal} {\bibinfo  {journal} {Phys. Rev. Lett.}\ }\textbf
  {\bibinfo {volume} {90}},\ \bibinfo {pages} {167903} (\bibinfo {year}
  {2003})}\BibitemShut {NoStop}%
\bibitem [{\citenamefont {Villar}\ \emph {et~al.}(2006)\citenamefont {Villar},
  \citenamefont {Martinelli}, \citenamefont {Fabre},\ and\ \citenamefont
  {Nussenzveig}}]{PhysRevLett.97.140504}%
  \BibitemOpen
  \bibfield  {author} {\bibinfo {author} {\bibfnamefont {A.~S.}\ \bibnamefont
  {Villar}}, \bibinfo {author} {\bibfnamefont {M.}~\bibnamefont {Martinelli}},
  \bibinfo {author} {\bibfnamefont {C.}~\bibnamefont {Fabre}}, \ and\ \bibinfo
  {author} {\bibfnamefont {P.}~\bibnamefont {Nussenzveig}},\ }\href {\doibase
  10.1103/PhysRevLett.97.140504} {\bibfield  {journal} {\bibinfo  {journal}
  {Phys. Rev. Lett.}\ }\textbf {\bibinfo {volume} {97}},\ \bibinfo {pages}
  {140504} (\bibinfo {year} {2006})}\BibitemShut {NoStop}%
\bibitem [{\citenamefont {D\"ur}\ \emph {et~al.}(2000)\citenamefont {D\"ur},
  \citenamefont {Vidal},\ and\ \citenamefont {Cirac}}]{PhysRevA.62.062314}%
  \BibitemOpen
  \bibfield  {author} {\bibinfo {author} {\bibfnamefont {W.}~\bibnamefont
  {D\"ur}}, \bibinfo {author} {\bibfnamefont {G.}~\bibnamefont {Vidal}}, \ and\
  \bibinfo {author} {\bibfnamefont {J.~I.}\ \bibnamefont {Cirac}},\ }\href
  {\doibase 10.1103/PhysRevA.62.062314} {\bibfield  {journal} {\bibinfo
  {journal} {Phys. Rev. A}\ }\textbf {\bibinfo {volume} {62}},\ \bibinfo
  {pages} {062314} (\bibinfo {year} {2000})}\BibitemShut {NoStop}%
\bibitem [{\citenamefont {Bennett}\ \emph {et~al.}(1999)\citenamefont
  {Bennett}, \citenamefont {DiVincenzo}, \citenamefont {Mor}, \citenamefont
  {Shor}, \citenamefont {Smolin},\ and\ \citenamefont
  {Terhal}}]{PhysRevLett.82.5385}%
  \BibitemOpen
  \bibfield  {author} {\bibinfo {author} {\bibfnamefont {C.~H.}\ \bibnamefont
  {Bennett}}, \bibinfo {author} {\bibfnamefont {D.~P.}\ \bibnamefont
  {DiVincenzo}}, \bibinfo {author} {\bibfnamefont {T.}~\bibnamefont {Mor}},
  \bibinfo {author} {\bibfnamefont {P.~W.}\ \bibnamefont {Shor}}, \bibinfo
  {author} {\bibfnamefont {J.~A.}\ \bibnamefont {Smolin}}, \ and\ \bibinfo
  {author} {\bibfnamefont {B.~M.}\ \bibnamefont {Terhal}},\ }\href {\doibase
  10.1103/PhysRevLett.82.5385} {\bibfield  {journal} {\bibinfo  {journal}
  {Phys. Rev. Lett.}\ }\textbf {\bibinfo {volume} {82}},\ \bibinfo {pages}
  {5385} (\bibinfo {year} {1999})}\BibitemShut {NoStop}%
\bibitem [{\citenamefont {Kustura}\ \emph {et~al.}(2022)\citenamefont
  {Kustura}, \citenamefont {Gonzalez-Ballestero}, \citenamefont {Sommer},
  \citenamefont {Meyer}, \citenamefont {Quidant},\ and\ \citenamefont
  {Romero-Isart}}]{PhysRevLett.128.143601}%
  \BibitemOpen
  \bibfield  {author} {\bibinfo {author} {\bibfnamefont {K.}~\bibnamefont
  {Kustura}}, \bibinfo {author} {\bibfnamefont {C.}~\bibnamefont
  {Gonzalez-Ballestero}}, \bibinfo {author} {\bibfnamefont {A.~d. l.~R.}\
  \bibnamefont {Sommer}}, \bibinfo {author} {\bibfnamefont {N.}~\bibnamefont
  {Meyer}}, \bibinfo {author} {\bibfnamefont {R.}~\bibnamefont {Quidant}}, \
  and\ \bibinfo {author} {\bibfnamefont {O.}~\bibnamefont {Romero-Isart}},\
  }\href {\doibase 10.1103/PhysRevLett.128.143601} {\bibfield  {journal}
  {\bibinfo  {journal} {Phys. Rev. Lett.}\ }\textbf {\bibinfo {volume} {128}},\
  \bibinfo {pages} {143601} (\bibinfo {year} {2022})}\BibitemShut {NoStop}%
\bibitem [{\citenamefont {Pontin}\ \emph {et~al.}(2014)\citenamefont {Pontin},
  \citenamefont {Bonaldi}, \citenamefont {Borrielli}, \citenamefont
  {Cataliotti}, \citenamefont {Marino}, \citenamefont {Prodi}, \citenamefont
  {Serra},\ and\ \citenamefont {Marin}}]{PhysRevLett.112.023601}%
  \BibitemOpen
  \bibfield  {author} {\bibinfo {author} {\bibfnamefont {A.}~\bibnamefont
  {Pontin}}, \bibinfo {author} {\bibfnamefont {M.}~\bibnamefont {Bonaldi}},
  \bibinfo {author} {\bibfnamefont {A.}~\bibnamefont {Borrielli}}, \bibinfo
  {author} {\bibfnamefont {F.~S.}\ \bibnamefont {Cataliotti}}, \bibinfo
  {author} {\bibfnamefont {F.}~\bibnamefont {Marino}}, \bibinfo {author}
  {\bibfnamefont {G.~A.}\ \bibnamefont {Prodi}}, \bibinfo {author}
  {\bibfnamefont {E.}~\bibnamefont {Serra}}, \ and\ \bibinfo {author}
  {\bibfnamefont {F.}~\bibnamefont {Marin}},\ }\href {\doibase
  10.1103/PhysRevLett.112.023601} {\bibfield  {journal} {\bibinfo  {journal}
  {Phys. Rev. Lett.}\ }\textbf {\bibinfo {volume} {112}},\ \bibinfo {pages}
  {023601} (\bibinfo {year} {2014})}\BibitemShut {NoStop}%
\bibitem [{\citenamefont {Kienzler}\ \emph {et~al.}(2016)\citenamefont
  {Kienzler}, \citenamefont {Fl\"uhmann}, \citenamefont {Negnevitsky},
  \citenamefont {Lo}, \citenamefont {Marinelli}, \citenamefont {Nadlinger},\
  and\ \citenamefont {Home}}]{PhysRevLett.116.140402}%
  \BibitemOpen
  \bibfield  {author} {\bibinfo {author} {\bibfnamefont {D.}~\bibnamefont
  {Kienzler}}, \bibinfo {author} {\bibfnamefont {C.}~\bibnamefont
  {Fl\"uhmann}}, \bibinfo {author} {\bibfnamefont {V.}~\bibnamefont
  {Negnevitsky}}, \bibinfo {author} {\bibfnamefont {H.-Y.}\ \bibnamefont {Lo}},
  \bibinfo {author} {\bibfnamefont {M.}~\bibnamefont {Marinelli}}, \bibinfo
  {author} {\bibfnamefont {D.}~\bibnamefont {Nadlinger}}, \ and\ \bibinfo
  {author} {\bibfnamefont {J.~P.}\ \bibnamefont {Home}},\ }\href {\doibase
  10.1103/PhysRevLett.116.140402} {\bibfield  {journal} {\bibinfo  {journal}
  {Phys. Rev. Lett.}\ }\textbf {\bibinfo {volume} {116}},\ \bibinfo {pages}
  {140402} (\bibinfo {year} {2016})}\BibitemShut {NoStop}%
\bibitem [{\citenamefont {Cirac}\ \emph {et~al.}(1997)\citenamefont {Cirac},
  \citenamefont {Zoller}, \citenamefont {Kimble},\ and\ \citenamefont
  {Mabuchi}}]{PhysRevLett.78.3221}%
  \BibitemOpen
  \bibfield  {author} {\bibinfo {author} {\bibfnamefont {J.~I.}\ \bibnamefont
  {Cirac}}, \bibinfo {author} {\bibfnamefont {P.}~\bibnamefont {Zoller}},
  \bibinfo {author} {\bibfnamefont {H.~J.}\ \bibnamefont {Kimble}}, \ and\
  \bibinfo {author} {\bibfnamefont {H.}~\bibnamefont {Mabuchi}},\ }\href
  {\doibase 10.1103/PhysRevLett.78.3221} {\bibfield  {journal} {\bibinfo
  {journal} {Phys. Rev. Lett.}\ }\textbf {\bibinfo {volume} {78}},\ \bibinfo
  {pages} {3221} (\bibinfo {year} {1997})}\BibitemShut {NoStop}%
\bibitem [{\citenamefont {Duan}\ and\ \citenamefont
  {Monroe}(2010)}]{RevModPhys.82.1209}%
  \BibitemOpen
  \bibfield  {author} {\bibinfo {author} {\bibfnamefont {L.-M.}\ \bibnamefont
  {Duan}}\ and\ \bibinfo {author} {\bibfnamefont {C.}~\bibnamefont {Monroe}},\
  }\href {\doibase 10.1103/RevModPhys.82.1209} {\bibfield  {journal} {\bibinfo
  {journal} {Rev. Mod. Phys.}\ }\textbf {\bibinfo {volume} {82}},\ \bibinfo
  {pages} {1209} (\bibinfo {year} {2010})}\BibitemShut {NoStop}%
\bibitem [{\citenamefont {Deng}\ \emph {et~al.}(2017)\citenamefont {Deng},
  \citenamefont {Li},\ and\ \citenamefont {Das~Sarma}}]{PhysRevX.7.021021}%
  \BibitemOpen
  \bibfield  {author} {\bibinfo {author} {\bibfnamefont {D.-L.}\ \bibnamefont
  {Deng}}, \bibinfo {author} {\bibfnamefont {X.}~\bibnamefont {Li}}, \ and\
  \bibinfo {author} {\bibfnamefont {S.}~\bibnamefont {Das~Sarma}},\ }\href
  {\doibase 10.1103/PhysRevX.7.021021} {\bibfield  {journal} {\bibinfo
  {journal} {Phys. Rev. X}\ }\textbf {\bibinfo {volume} {7}},\ \bibinfo {pages}
  {021021} (\bibinfo {year} {2017})}\BibitemShut {NoStop}%
\bibitem [{\citenamefont {Cotrufo}\ \emph
  {et~al.}(2017{\natexlab{a}})\citenamefont {Cotrufo}, \citenamefont {Fiore},\
  and\ \citenamefont {Verhagen}}]{PhysRevLett.118.133603}%
  \BibitemOpen
  \bibfield  {author} {\bibinfo {author} {\bibfnamefont {M.}~\bibnamefont
  {Cotrufo}}, \bibinfo {author} {\bibfnamefont {A.}~\bibnamefont {Fiore}}, \
  and\ \bibinfo {author} {\bibfnamefont {E.}~\bibnamefont {Verhagen}},\ }\href
  {\doibase 10.1103/PhysRevLett.118.133603} {\bibfield  {journal} {\bibinfo
  {journal} {Phys. Rev. Lett.}\ }\textbf {\bibinfo {volume} {118}},\ \bibinfo
  {pages} {133603} (\bibinfo {year} {2017}{\natexlab{a}})}\BibitemShut
  {NoStop}%
\bibitem [{\citenamefont {Zhou}\ \emph {et~al.}(2022)\citenamefont {Zhou},
  \citenamefont {Hu}, \citenamefont {L{\"u}}, \citenamefont {Li}, \citenamefont
  {Huang}, \citenamefont {Xiong},\ and\ \citenamefont
  {L{\"u}}}]{zhou2022synergistic}%
  \BibitemOpen
  \bibfield  {author} {\bibinfo {author} {\bibfnamefont {Y.}~\bibnamefont
  {Zhou}}, \bibinfo {author} {\bibfnamefont {C.-S.}\ \bibnamefont {Hu}},
  \bibinfo {author} {\bibfnamefont {D.-Y.}\ \bibnamefont {L{\"u}}}, \bibinfo
  {author} {\bibfnamefont {X.-K.}\ \bibnamefont {Li}}, \bibinfo {author}
  {\bibfnamefont {H.-M.}\ \bibnamefont {Huang}}, \bibinfo {author}
  {\bibfnamefont {Y.-C.}\ \bibnamefont {Xiong}}, \ and\ \bibinfo {author}
  {\bibfnamefont {X.-Y.}\ \bibnamefont {L{\"u}}},\ }\href@noop {} {\bibfield
  {journal} {\bibinfo  {journal} {Photonics Research}\ }\textbf {\bibinfo
  {volume} {10}},\ \bibinfo {pages} {1640} (\bibinfo {year}
  {2022})}\BibitemShut {NoStop}%
\bibitem [{\citenamefont {Hei}\ \emph {et~al.}(2023{\natexlab{a}})\citenamefont
  {Hei}, \citenamefont {Li}, \citenamefont {Pan},\ and\ \citenamefont
  {Nori}}]{PhysRevLett.130.073602}%
  \BibitemOpen
  \bibfield  {author} {\bibinfo {author} {\bibfnamefont {X.-L.}\ \bibnamefont
  {Hei}}, \bibinfo {author} {\bibfnamefont {P.-B.}\ \bibnamefont {Li}},
  \bibinfo {author} {\bibfnamefont {X.-F.}\ \bibnamefont {Pan}}, \ and\
  \bibinfo {author} {\bibfnamefont {F.}~\bibnamefont {Nori}},\ }\href {\doibase
  10.1103/PhysRevLett.130.073602} {\bibfield  {journal} {\bibinfo  {journal}
  {Phys. Rev. Lett.}\ }\textbf {\bibinfo {volume} {130}},\ \bibinfo {pages}
  {073602} (\bibinfo {year} {2023}{\natexlab{a}})}\BibitemShut {NoStop}%
\bibitem [{\citenamefont {Birnbaum}\ \emph {et~al.}(2005)\citenamefont
  {Birnbaum}, \citenamefont {Boca}, \citenamefont {Miller}, \citenamefont
  {Boozer}, \citenamefont {Northup},\ and\ \citenamefont
  {Kimble}}]{birnbaum2005photon}%
  \BibitemOpen
  \bibfield  {author} {\bibinfo {author} {\bibfnamefont {K.~M.}\ \bibnamefont
  {Birnbaum}}, \bibinfo {author} {\bibfnamefont {A.}~\bibnamefont {Boca}},
  \bibinfo {author} {\bibfnamefont {R.}~\bibnamefont {Miller}}, \bibinfo
  {author} {\bibfnamefont {A.~D.}\ \bibnamefont {Boozer}}, \bibinfo {author}
  {\bibfnamefont {T.~E.}\ \bibnamefont {Northup}}, \ and\ \bibinfo {author}
  {\bibfnamefont {H.~J.}\ \bibnamefont {Kimble}},\ }\href@noop {} {\bibfield
  {journal} {\bibinfo  {journal} {Nature}\ }\textbf {\bibinfo {volume} {436}},\
  \bibinfo {pages} {87} (\bibinfo {year} {2005})}\BibitemShut {NoStop}%
\bibitem [{\citenamefont {M{\"u}cke}\ \emph {et~al.}(2010)\citenamefont
  {M{\"u}cke}, \citenamefont {Figueroa}, \citenamefont {Bochmann},
  \citenamefont {Hahn}, \citenamefont {Murr}, \citenamefont {Ritter},
  \citenamefont {Villas-Boas},\ and\ \citenamefont
  {Rempe}}]{mucke2010electromagnetically}%
  \BibitemOpen
  \bibfield  {author} {\bibinfo {author} {\bibfnamefont {M.}~\bibnamefont
  {M{\"u}cke}}, \bibinfo {author} {\bibfnamefont {E.}~\bibnamefont {Figueroa}},
  \bibinfo {author} {\bibfnamefont {J.}~\bibnamefont {Bochmann}}, \bibinfo
  {author} {\bibfnamefont {C.}~\bibnamefont {Hahn}}, \bibinfo {author}
  {\bibfnamefont {K.}~\bibnamefont {Murr}}, \bibinfo {author} {\bibfnamefont
  {S.}~\bibnamefont {Ritter}}, \bibinfo {author} {\bibfnamefont {C.~J.}\
  \bibnamefont {Villas-Boas}}, \ and\ \bibinfo {author} {\bibfnamefont
  {G.}~\bibnamefont {Rempe}},\ }\href@noop {} {\bibfield  {journal} {\bibinfo
  {journal} {Nature}\ }\textbf {\bibinfo {volume} {465}},\ \bibinfo {pages}
  {755} (\bibinfo {year} {2010})}\BibitemShut {NoStop}%
\bibitem [{\citenamefont {Hamsen}\ \emph {et~al.}(2017)\citenamefont {Hamsen},
  \citenamefont {Tolazzi}, \citenamefont {Wilk},\ and\ \citenamefont
  {Rempe}}]{PhysRevLett.118.133604}%
  \BibitemOpen
  \bibfield  {author} {\bibinfo {author} {\bibfnamefont {C.}~\bibnamefont
  {Hamsen}}, \bibinfo {author} {\bibfnamefont {K.~N.}\ \bibnamefont {Tolazzi}},
  \bibinfo {author} {\bibfnamefont {T.}~\bibnamefont {Wilk}}, \ and\ \bibinfo
  {author} {\bibfnamefont {G.}~\bibnamefont {Rempe}},\ }\href {\doibase
  10.1103/PhysRevLett.118.133604} {\bibfield  {journal} {\bibinfo  {journal}
  {Phys. Rev. Lett.}\ }\textbf {\bibinfo {volume} {118}},\ \bibinfo {pages}
  {133604} (\bibinfo {year} {2017})}\BibitemShut {NoStop}%
\bibitem [{\citenamefont {Lecocq}\ \emph {et~al.}(2015)\citenamefont {Lecocq},
  \citenamefont {Teufel}, \citenamefont {Aumentado},\ and\ \citenamefont
  {Simmonds}}]{lecocq2015resolving}%
  \BibitemOpen
  \bibfield  {author} {\bibinfo {author} {\bibfnamefont {F.}~\bibnamefont
  {Lecocq}}, \bibinfo {author} {\bibfnamefont {J.~D.}\ \bibnamefont {Teufel}},
  \bibinfo {author} {\bibfnamefont {J.}~\bibnamefont {Aumentado}}, \ and\
  \bibinfo {author} {\bibfnamefont {R.~W.}\ \bibnamefont {Simmonds}},\
  }\href@noop {} {\bibfield  {journal} {\bibinfo  {journal} {Nature Physics}\
  }\textbf {\bibinfo {volume} {11}},\ \bibinfo {pages} {635} (\bibinfo {year}
  {2015})}\BibitemShut {NoStop}%
\bibitem [{\citenamefont {Wolf}\ \emph {et~al.}(2019)\citenamefont {Wolf},
  \citenamefont {Shi}, \citenamefont {Heip}, \citenamefont {Gessner},
  \citenamefont {Pezz{\`e}}, \citenamefont {Smerzi}, \citenamefont {Schulte},
  \citenamefont {Hammerer},\ and\ \citenamefont {Schmidt}}]{wolf2019motional}%
  \BibitemOpen
  \bibfield  {author} {\bibinfo {author} {\bibfnamefont {F.}~\bibnamefont
  {Wolf}}, \bibinfo {author} {\bibfnamefont {C.}~\bibnamefont {Shi}}, \bibinfo
  {author} {\bibfnamefont {J.~C.}\ \bibnamefont {Heip}}, \bibinfo {author}
  {\bibfnamefont {M.}~\bibnamefont {Gessner}}, \bibinfo {author} {\bibfnamefont
  {L.}~\bibnamefont {Pezz{\`e}}}, \bibinfo {author} {\bibfnamefont
  {A.}~\bibnamefont {Smerzi}}, \bibinfo {author} {\bibfnamefont
  {M.}~\bibnamefont {Schulte}}, \bibinfo {author} {\bibfnamefont
  {K.}~\bibnamefont {Hammerer}}, \ and\ \bibinfo {author} {\bibfnamefont
  {P.~O.}\ \bibnamefont {Schmidt}},\ }\href@noop {} {\bibfield  {journal}
  {\bibinfo  {journal} {Nature communications}\ }\textbf {\bibinfo {volume}
  {10}},\ \bibinfo {pages} {2929} (\bibinfo {year} {2019})}\BibitemShut
  {NoStop}%
\bibitem [{\citenamefont {Ma}\ \emph {et~al.}(2021)\citenamefont {Ma},
  \citenamefont {Viennot}, \citenamefont {Kotler}, \citenamefont {Teufel},\
  and\ \citenamefont {Lehnert}}]{ma2021non}%
  \BibitemOpen
  \bibfield  {author} {\bibinfo {author} {\bibfnamefont {X.}~\bibnamefont
  {Ma}}, \bibinfo {author} {\bibfnamefont {J.~J.}\ \bibnamefont {Viennot}},
  \bibinfo {author} {\bibfnamefont {S.}~\bibnamefont {Kotler}}, \bibinfo
  {author} {\bibfnamefont {J.~D.}\ \bibnamefont {Teufel}}, \ and\ \bibinfo
  {author} {\bibfnamefont {K.~W.}\ \bibnamefont {Lehnert}},\ }\href@noop {}
  {\bibfield  {journal} {\bibinfo  {journal} {Nature Physics}\ }\textbf
  {\bibinfo {volume} {17}},\ \bibinfo {pages} {322} (\bibinfo {year}
  {2021})}\BibitemShut {NoStop}%
\bibitem [{\citenamefont {Kozlov}\ \emph {et~al.}(2018)\citenamefont {Kozlov},
  \citenamefont {Safronova}, \citenamefont {Crespo L\'opez-Urrutia},\ and\
  \citenamefont {Schmidt}}]{RevModPhys.90.045005}%
  \BibitemOpen
  \bibfield  {author} {\bibinfo {author} {\bibfnamefont {M.~G.}\ \bibnamefont
  {Kozlov}}, \bibinfo {author} {\bibfnamefont {M.~S.}\ \bibnamefont
  {Safronova}}, \bibinfo {author} {\bibfnamefont {J.~R.}\ \bibnamefont {Crespo
  L\'opez-Urrutia}}, \ and\ \bibinfo {author} {\bibfnamefont {P.~O.}\
  \bibnamefont {Schmidt}},\ }\href {\doibase 10.1103/RevModPhys.90.045005}
  {\bibfield  {journal} {\bibinfo  {journal} {Rev. Mod. Phys.}\ }\textbf
  {\bibinfo {volume} {90}},\ \bibinfo {pages} {045005} (\bibinfo {year}
  {2018})}\BibitemShut {NoStop}%
\bibitem [{\citenamefont {Pikovski}\ \emph {et~al.}(2012)\citenamefont
  {Pikovski}, \citenamefont {Vanner}, \citenamefont {Aspelmeyer}, \citenamefont
  {Kim},\ and\ \citenamefont {Brukner}}]{pikovski2012probing}%
  \BibitemOpen
  \bibfield  {author} {\bibinfo {author} {\bibfnamefont {I.}~\bibnamefont
  {Pikovski}}, \bibinfo {author} {\bibfnamefont {M.~R.}\ \bibnamefont
  {Vanner}}, \bibinfo {author} {\bibfnamefont {M.}~\bibnamefont {Aspelmeyer}},
  \bibinfo {author} {\bibfnamefont {M.}~\bibnamefont {Kim}}, \ and\ \bibinfo
  {author} {\bibfnamefont {{\v{C}}.}~\bibnamefont {Brukner}},\ }\href@noop {}
  {\bibfield  {journal} {\bibinfo  {journal} {Nature Physics}\ }\textbf
  {\bibinfo {volume} {8}},\ \bibinfo {pages} {393} (\bibinfo {year}
  {2012})}\BibitemShut {NoStop}%
\bibitem [{\citenamefont {Shomroni}\ \emph {et~al.}(2019)\citenamefont
  {Shomroni}, \citenamefont {Youssefi}, \citenamefont {Sauerwein},
  \citenamefont {Qiu}, \citenamefont {Seidler}, \citenamefont {Malz},
  \citenamefont {Nunnenkamp},\ and\ \citenamefont
  {Kippenberg}}]{PhysRevX.9.041022}%
  \BibitemOpen
  \bibfield  {author} {\bibinfo {author} {\bibfnamefont {I.}~\bibnamefont
  {Shomroni}}, \bibinfo {author} {\bibfnamefont {A.}~\bibnamefont {Youssefi}},
  \bibinfo {author} {\bibfnamefont {N.}~\bibnamefont {Sauerwein}}, \bibinfo
  {author} {\bibfnamefont {L.}~\bibnamefont {Qiu}}, \bibinfo {author}
  {\bibfnamefont {P.}~\bibnamefont {Seidler}}, \bibinfo {author} {\bibfnamefont
  {D.}~\bibnamefont {Malz}}, \bibinfo {author} {\bibfnamefont {A.}~\bibnamefont
  {Nunnenkamp}}, \ and\ \bibinfo {author} {\bibfnamefont {T.~J.}\ \bibnamefont
  {Kippenberg}},\ }\href {\doibase 10.1103/PhysRevX.9.041022} {\bibfield
  {journal} {\bibinfo  {journal} {Phys. Rev. X}\ }\textbf {\bibinfo {volume}
  {9}},\ \bibinfo {pages} {041022} (\bibinfo {year} {2019})}\BibitemShut
  {NoStop}%
\bibitem [{\citenamefont {Vogel}\ and\ \citenamefont
  {Filho}(1995)}]{PhysRevA.52.4214}%
  \BibitemOpen
  \bibfield  {author} {\bibinfo {author} {\bibfnamefont {W.}~\bibnamefont
  {Vogel}}\ and\ \bibinfo {author} {\bibfnamefont {R.~L. d.~M.}\ \bibnamefont
  {Filho}},\ }\href {\doibase 10.1103/PhysRevA.52.4214} {\bibfield  {journal}
  {\bibinfo  {journal} {Phys. Rev. A}\ }\textbf {\bibinfo {volume} {52}},\
  \bibinfo {pages} {4214} (\bibinfo {year} {1995})}\BibitemShut {NoStop}%
\bibitem [{SM()}]{SM}%
  \BibitemOpen
  \href@noop {} {\bibinfo  {journal} {See Supplemental Material for more
  detailed account of (I) derivation of the tripartite Hamiltonian and
  calculating the energy spectrum, (II) numerical solving the master equation,
  (III) resolving Vacuum Fluctuations from Schr\"{o}dinger picture dynamics,
  and (IV) studying the effect of motional phonon decay.}\ }\BibitemShut
  {NoStop}%
\bibitem [{\citenamefont {Lai}\ \emph {et~al.}(2022)\citenamefont {Lai},
  \citenamefont {Chen}, \citenamefont {Qin}, \citenamefont {Miranowicz},\ and\
  \citenamefont {Nori}}]{lai2022tripartite}%
  \BibitemOpen
\bibfield  {journal} {  }\bibfield  {author} {\bibinfo {author} {\bibfnamefont
  {D.-G.}\ \bibnamefont {Lai}}, \bibinfo {author} {\bibfnamefont {Y.-H.}\
  \bibnamefont {Chen}}, \bibinfo {author} {\bibfnamefont {W.}~\bibnamefont
  {Qin}}, \bibinfo {author} {\bibfnamefont {A.}~\bibnamefont {Miranowicz}}, \
  and\ \bibinfo {author} {\bibfnamefont {F.}~\bibnamefont {Nori}},\ }\href@noop
  {} {\bibfield  {journal} {\bibinfo  {journal} {Physical Review Research}\
  }\textbf {\bibinfo {volume} {4}},\ \bibinfo {pages} {033112} (\bibinfo {year}
  {2022})}\BibitemShut {NoStop}%
\bibitem [{\citenamefont {Cotrufo}\ \emph
  {et~al.}(2017{\natexlab{b}})\citenamefont {Cotrufo}, \citenamefont {Fiore},\
  and\ \citenamefont {Verhagen}}]{cotrufo2017coherent}%
  \BibitemOpen
  \bibfield  {author} {\bibinfo {author} {\bibfnamefont {M.}~\bibnamefont
  {Cotrufo}}, \bibinfo {author} {\bibfnamefont {A.}~\bibnamefont {Fiore}}, \
  and\ \bibinfo {author} {\bibfnamefont {E.}~\bibnamefont {Verhagen}},\
  }\href@noop {} {\bibfield  {journal} {\bibinfo  {journal} {Physical Review
  Letters}\ }\textbf {\bibinfo {volume} {118}},\ \bibinfo {pages} {133603}
  (\bibinfo {year} {2017}{\natexlab{b}})}\BibitemShut {NoStop}%
\bibitem [{\citenamefont {Hei}\ \emph {et~al.}(2023{\natexlab{b}})\citenamefont
  {Hei}, \citenamefont {Li}, \citenamefont {Pan},\ and\ \citenamefont
  {Nori}}]{hei2023enhanced}%
  \BibitemOpen
  \bibfield  {author} {\bibinfo {author} {\bibfnamefont {X.-L.}\ \bibnamefont
  {Hei}}, \bibinfo {author} {\bibfnamefont {P.-B.}\ \bibnamefont {Li}},
  \bibinfo {author} {\bibfnamefont {X.-F.}\ \bibnamefont {Pan}}, \ and\
  \bibinfo {author} {\bibfnamefont {F.}~\bibnamefont {Nori}},\ }\href@noop {}
  {\bibfield  {journal} {\bibinfo  {journal} {Physical Review Letters}\
  }\textbf {\bibinfo {volume} {130}},\ \bibinfo {pages} {073602} (\bibinfo
  {year} {2023}{\natexlab{b}})}\BibitemShut {NoStop}%
\bibitem [{\citenamefont {Peng}\ and\ \citenamefont
  {Deng}(2022)}]{PhysRevA.105.043711}%
  \BibitemOpen
  \bibfield  {author} {\bibinfo {author} {\bibfnamefont {Z.}~\bibnamefont
  {Peng}}\ and\ \bibinfo {author} {\bibfnamefont {Y.}~\bibnamefont {Deng}},\
  }\href {\doibase 10.1103/PhysRevA.105.043711} {\bibfield  {journal} {\bibinfo
   {journal} {Phys. Rev. A}\ }\textbf {\bibinfo {volume} {105}},\ \bibinfo
  {pages} {043711} (\bibinfo {year} {2022})}\BibitemShut {NoStop}%
\bibitem [{\citenamefont {Ludlow}\ \emph {et~al.}(2015)\citenamefont {Ludlow},
  \citenamefont {Boyd}, \citenamefont {Ye}, \citenamefont {Peik},\ and\
  \citenamefont {Schmidt}}]{RevModPhys.87.637}%
  \BibitemOpen
  \bibfield  {author} {\bibinfo {author} {\bibfnamefont {A.~D.}\ \bibnamefont
  {Ludlow}}, \bibinfo {author} {\bibfnamefont {M.~M.}\ \bibnamefont {Boyd}},
  \bibinfo {author} {\bibfnamefont {J.}~\bibnamefont {Ye}}, \bibinfo {author}
  {\bibfnamefont {E.}~\bibnamefont {Peik}}, \ and\ \bibinfo {author}
  {\bibfnamefont {P.~O.}\ \bibnamefont {Schmidt}},\ }\href {\doibase
  10.1103/RevModPhys.87.637} {\bibfield  {journal} {\bibinfo  {journal} {Rev.
  Mod. Phys.}\ }\textbf {\bibinfo {volume} {87}},\ \bibinfo {pages} {637}
  (\bibinfo {year} {2015})}\BibitemShut {NoStop}%
\bibitem [{\citenamefont {Norcia}\ and\ \citenamefont
  {Thompson}(2016)}]{PhysRevX.6.011025}%
  \BibitemOpen
  \bibfield  {author} {\bibinfo {author} {\bibfnamefont {M.~A.}\ \bibnamefont
  {Norcia}}\ and\ \bibinfo {author} {\bibfnamefont {J.~K.}\ \bibnamefont
  {Thompson}},\ }\href {\doibase 10.1103/PhysRevX.6.011025} {\bibfield
  {journal} {\bibinfo  {journal} {Phys. Rev. X}\ }\textbf {\bibinfo {volume}
  {6}},\ \bibinfo {pages} {011025} (\bibinfo {year} {2016})}\BibitemShut
  {NoStop}%
\bibitem [{\citenamefont {Winchester}\ \emph {et~al.}(2017)\citenamefont
  {Winchester}, \citenamefont {Norcia}, \citenamefont {Cline},\ and\
  \citenamefont {Thompson}}]{PhysRevLett.118.263601}%
  \BibitemOpen
  \bibfield  {author} {\bibinfo {author} {\bibfnamefont {M.~N.}\ \bibnamefont
  {Winchester}}, \bibinfo {author} {\bibfnamefont {M.~A.}\ \bibnamefont
  {Norcia}}, \bibinfo {author} {\bibfnamefont {J.~R.~K.}\ \bibnamefont
  {Cline}}, \ and\ \bibinfo {author} {\bibfnamefont {J.~K.}\ \bibnamefont
  {Thompson}},\ }\href {\doibase 10.1103/PhysRevLett.118.263601} {\bibfield
  {journal} {\bibinfo  {journal} {Phys. Rev. Lett.}\ }\textbf {\bibinfo
  {volume} {118}},\ \bibinfo {pages} {263601} (\bibinfo {year}
  {2017})}\BibitemShut {NoStop}%
\bibitem [{\citenamefont {Wolke}\ \emph {et~al.}(2012)\citenamefont {Wolke},
  \citenamefont {Klinner}, \citenamefont {Ke{\ss}ler},\ and\ \citenamefont
  {Hemmerich}}]{wolke2012cavity}%
  \BibitemOpen
  \bibfield  {author} {\bibinfo {author} {\bibfnamefont {M.}~\bibnamefont
  {Wolke}}, \bibinfo {author} {\bibfnamefont {J.}~\bibnamefont {Klinner}},
  \bibinfo {author} {\bibfnamefont {H.}~\bibnamefont {Ke{\ss}ler}}, \ and\
  \bibinfo {author} {\bibfnamefont {A.}~\bibnamefont {Hemmerich}},\ }\href@noop
  {} {\bibfield  {journal} {\bibinfo  {journal} {Science}\ }\textbf {\bibinfo
  {volume} {337}},\ \bibinfo {pages} {75} (\bibinfo {year} {2012})}\BibitemShut
  {NoStop}%
\bibitem [{\citenamefont {L{\'e}onard}\ \emph
  {et~al.}(2017{\natexlab{b}})\citenamefont {L{\'e}onard}, \citenamefont
  {Morales}, \citenamefont {Zupancic}, \citenamefont {Donner},\ and\
  \citenamefont {Esslinger}}]{leonard2017monitoring}%
  \BibitemOpen
  \bibfield  {author} {\bibinfo {author} {\bibfnamefont {J.}~\bibnamefont
  {L{\'e}onard}}, \bibinfo {author} {\bibfnamefont {A.}~\bibnamefont
  {Morales}}, \bibinfo {author} {\bibfnamefont {P.}~\bibnamefont {Zupancic}},
  \bibinfo {author} {\bibfnamefont {T.}~\bibnamefont {Donner}}, \ and\ \bibinfo
  {author} {\bibfnamefont {T.}~\bibnamefont {Esslinger}},\ }\href@noop {}
  {\bibfield  {journal} {\bibinfo  {journal} {Science}\ }\textbf {\bibinfo
  {volume} {358}},\ \bibinfo {pages} {1415} (\bibinfo {year}
  {2017}{\natexlab{b}})}\BibitemShut {NoStop}%
\bibitem [{\citenamefont {Royer}(1977)}]{PhysRevA.15.449}%
  \BibitemOpen
  \bibfield  {author} {\bibinfo {author} {\bibfnamefont {A.}~\bibnamefont
  {Royer}},\ }\href {\doibase 10.1103/PhysRevA.15.449} {\bibfield  {journal}
  {\bibinfo  {journal} {Phys. Rev. A}\ }\textbf {\bibinfo {volume} {15}},\
  \bibinfo {pages} {449} (\bibinfo {year} {1977})}\BibitemShut {NoStop}%
\bibitem [{\citenamefont {Bohmann}\ and\ \citenamefont
  {Agudelo}(2020)}]{PhysRevLett.124.133601}%
  \BibitemOpen
  \bibfield  {author} {\bibinfo {author} {\bibfnamefont {M.}~\bibnamefont
  {Bohmann}}\ and\ \bibinfo {author} {\bibfnamefont {E.}~\bibnamefont
  {Agudelo}},\ }\href {\doibase 10.1103/PhysRevLett.124.133601} {\bibfield
  {journal} {\bibinfo  {journal} {Phys. Rev. Lett.}\ }\textbf {\bibinfo
  {volume} {124}},\ \bibinfo {pages} {133601} (\bibinfo {year}
  {2020})}\BibitemShut {NoStop}%
\bibitem [{\citenamefont {Biagi}\ \emph {et~al.}(2021)\citenamefont {Biagi},
  \citenamefont {Bohmann}, \citenamefont {Agudelo}, \citenamefont {Bellini},\
  and\ \citenamefont {Zavatta}}]{PhysRevLett.126.023605}%
  \BibitemOpen
  \bibfield  {author} {\bibinfo {author} {\bibfnamefont {N.}~\bibnamefont
  {Biagi}}, \bibinfo {author} {\bibfnamefont {M.}~\bibnamefont {Bohmann}},
  \bibinfo {author} {\bibfnamefont {E.}~\bibnamefont {Agudelo}}, \bibinfo
  {author} {\bibfnamefont {M.}~\bibnamefont {Bellini}}, \ and\ \bibinfo
  {author} {\bibfnamefont {A.}~\bibnamefont {Zavatta}},\ }\href {\doibase
  10.1103/PhysRevLett.126.023605} {\bibfield  {journal} {\bibinfo  {journal}
  {Phys. Rev. Lett.}\ }\textbf {\bibinfo {volume} {126}},\ \bibinfo {pages}
  {023605} (\bibinfo {year} {2021})}\BibitemShut {NoStop}%
\bibitem [{\citenamefont {Arndt}\ and\ \citenamefont
  {Hornberger}(2014)}]{arndt2014testing}%
  \BibitemOpen
  \bibfield  {author} {\bibinfo {author} {\bibfnamefont {M.}~\bibnamefont
  {Arndt}}\ and\ \bibinfo {author} {\bibfnamefont {K.}~\bibnamefont
  {Hornberger}},\ }\href@noop {} {\bibfield  {journal} {\bibinfo  {journal}
  {Nature Physics}\ }\textbf {\bibinfo {volume} {10}},\ \bibinfo {pages} {271}
  (\bibinfo {year} {2014})}\BibitemShut {NoStop}%
\bibitem [{\citenamefont {Monroe}\ \emph {et~al.}(2014)\citenamefont {Monroe},
  \citenamefont {Raussendorf}, \citenamefont {Ruthven}, \citenamefont {Brown},
  \citenamefont {Maunz}, \citenamefont {Duan},\ and\ \citenamefont
  {Kim}}]{PhysRevA.89.022317}%
  \BibitemOpen
  \bibfield  {author} {\bibinfo {author} {\bibfnamefont {C.}~\bibnamefont
  {Monroe}}, \bibinfo {author} {\bibfnamefont {R.}~\bibnamefont {Raussendorf}},
  \bibinfo {author} {\bibfnamefont {A.}~\bibnamefont {Ruthven}}, \bibinfo
  {author} {\bibfnamefont {K.~R.}\ \bibnamefont {Brown}}, \bibinfo {author}
  {\bibfnamefont {P.}~\bibnamefont {Maunz}}, \bibinfo {author} {\bibfnamefont
  {L.-M.}\ \bibnamefont {Duan}}, \ and\ \bibinfo {author} {\bibfnamefont
  {J.}~\bibnamefont {Kim}},\ }\href {\doibase 10.1103/PhysRevA.89.022317}
  {\bibfield  {journal} {\bibinfo  {journal} {Phys. Rev. A}\ }\textbf {\bibinfo
  {volume} {89}},\ \bibinfo {pages} {022317} (\bibinfo {year}
  {2014})}\BibitemShut {NoStop}%
\bibitem [{\citenamefont {Fitzsimons}(2017)}]{fitzsimons2017private}%
  \BibitemOpen
  \bibfield  {author} {\bibinfo {author} {\bibfnamefont {J.~F.}\ \bibnamefont
  {Fitzsimons}},\ }\href@noop {} {\bibfield  {journal} {\bibinfo  {journal}
  {npj Quantum Information}\ }\textbf {\bibinfo {volume} {3}},\ \bibinfo
  {pages} {23} (\bibinfo {year} {2017})}\BibitemShut {NoStop}%
\bibitem [{\citenamefont {Buhrman}\ \emph {et~al.}(2010)\citenamefont
  {Buhrman}, \citenamefont {Cleve}, \citenamefont {Massar},\ and\ \citenamefont
  {de~Wolf}}]{RevModPhys.82.665}%
  \BibitemOpen
  \bibfield  {author} {\bibinfo {author} {\bibfnamefont {H.}~\bibnamefont
  {Buhrman}}, \bibinfo {author} {\bibfnamefont {R.}~\bibnamefont {Cleve}},
  \bibinfo {author} {\bibfnamefont {S.}~\bibnamefont {Massar}}, \ and\ \bibinfo
  {author} {\bibfnamefont {R.}~\bibnamefont {de~Wolf}},\ }\href {\doibase
  10.1103/RevModPhys.82.665} {\bibfield  {journal} {\bibinfo  {journal} {Rev.
  Mod. Phys.}\ }\textbf {\bibinfo {volume} {82}},\ \bibinfo {pages} {665}
  (\bibinfo {year} {2010})}\BibitemShut {NoStop}%
\bibitem [{\citenamefont {Tan}(1999)}]{tan1999computational}%
  \BibitemOpen
  \bibfield  {author} {\bibinfo {author} {\bibfnamefont {S.~M.}\ \bibnamefont
  {Tan}},\ }\href@noop {} {\bibfield  {journal} {\bibinfo  {journal} {Journal
  of Optics B: Quantum and Semiclassical Optics}\ }\textbf {\bibinfo {volume}
  {1}},\ \bibinfo {pages} {424} (\bibinfo {year} {1999})}\BibitemShut {NoStop}%
\bibitem [{\citenamefont {Beaudoin}\ \emph {et~al.}(2011)\citenamefont
  {Beaudoin}, \citenamefont {Gambetta},\ and\ \citenamefont
  {Blais}}]{PhysRevA.84.043832}%
  \BibitemOpen
  \bibfield  {author} {\bibinfo {author} {\bibfnamefont {F.}~\bibnamefont
  {Beaudoin}}, \bibinfo {author} {\bibfnamefont {J.~M.}\ \bibnamefont
  {Gambetta}}, \ and\ \bibinfo {author} {\bibfnamefont {A.}~\bibnamefont
  {Blais}},\ }\href {\doibase 10.1103/PhysRevA.84.043832} {\bibfield  {journal}
  {\bibinfo  {journal} {Phys. Rev. A}\ }\textbf {\bibinfo {volume} {84}},\
  \bibinfo {pages} {043832} (\bibinfo {year} {2011})}\BibitemShut {NoStop}%
\bibitem [{\citenamefont {Ridolfo}\ \emph {et~al.}(2012)\citenamefont
  {Ridolfo}, \citenamefont {Leib}, \citenamefont {Savasta},\ and\ \citenamefont
  {Hartmann}}]{PhysRevLett.109.193602}%
  \BibitemOpen
  \bibfield  {author} {\bibinfo {author} {\bibfnamefont {A.}~\bibnamefont
  {Ridolfo}}, \bibinfo {author} {\bibfnamefont {M.}~\bibnamefont {Leib}},
  \bibinfo {author} {\bibfnamefont {S.}~\bibnamefont {Savasta}}, \ and\
  \bibinfo {author} {\bibfnamefont {M.~J.}\ \bibnamefont {Hartmann}},\ }\href
  {\doibase 10.1103/PhysRevLett.109.193602} {\bibfield  {journal} {\bibinfo
  {journal} {Phys. Rev. Lett.}\ }\textbf {\bibinfo {volume} {109}},\ \bibinfo
  {pages} {193602} (\bibinfo {year} {2012})}\BibitemShut {NoStop}%
\bibitem [{\citenamefont {Glauber}(1963)}]{PhysRev.130.2529}%
  \BibitemOpen
  \bibfield  {author} {\bibinfo {author} {\bibfnamefont {R.~J.}\ \bibnamefont
  {Glauber}},\ }\href {\doibase 10.1103/PhysRev.130.2529} {\bibfield  {journal}
  {\bibinfo  {journal} {Phys. Rev.}\ }\textbf {\bibinfo {volume} {130}},\
  \bibinfo {pages} {2529} (\bibinfo {year} {1963})}\BibitemShut {NoStop}%
\bibitem [{\citenamefont {Senellart}\ \emph {et~al.}(2017)\citenamefont
  {Senellart}, \citenamefont {Solomon},\ and\ \citenamefont
  {White}}]{senellart2017high}%
  \BibitemOpen
  \bibfield  {author} {\bibinfo {author} {\bibfnamefont {P.}~\bibnamefont
  {Senellart}}, \bibinfo {author} {\bibfnamefont {G.}~\bibnamefont {Solomon}},
  \ and\ \bibinfo {author} {\bibfnamefont {A.}~\bibnamefont {White}},\
  }\href@noop {} {\bibfield  {journal} {\bibinfo  {journal} {Nature
  nanotechnology}\ }\textbf {\bibinfo {volume} {12}},\ \bibinfo {pages} {1026}
  (\bibinfo {year} {2017})}\BibitemShut {NoStop}%
\end{thebibliography}

%merlin.mbs apsrev4-1.bst 2010-07-25 4.21a (PWD, AO, DPC) hacked
%Control: key (0)
%Control: author (72) initials jnrlst
%Control: editor formatted (1) identically to author
%Control: production of article title (-1) disabled
%Control: page (0) single
%Control: year (1) truncated
%Control: production of eprint (0) enabled
%

\setcounter{figure}{0}
\setcounter{equation}{0}
\setcounter{section}{0}

\clearpage
\onecolumngrid
\vspace{\columnsep}

\newcolumntype{Y}{>{\centering\arraybackslash}X}
\newcolumntype{Z}{>{\raggedleft\arraybackslash}X}

\newlength{\figwidth}
\setlength{\figwidth}{0.45\textwidth}

\renewcommand{\thefigure}{S\arabic{figure}}
\renewcommand{\theHfigure}{Supplement.\thefigure}
\renewcommand{\theequation}{S\arabic{equation}}
\renewcommand{\thesection}{\arabic{section}}

\begin{center}
	\large{\textbf{Supplemental Material:\\ Unveiling Vacuum Fluctuations and Nonclassical States with Cavity-Enhanced Tripartite Interactions}}\\~\\

	\normalfont{Jing Tang$^1$ and Yuangang Deng$^{1}$\\
	\textit{$^1$Guangdong Provincial Key Laboratory of Quantum Metrology and Sensing $\&$ School of Physics and Astronomy, Sun Yat-Sen University (Zhuhai Campus), Zhuhai 519082, China
	}%
}

\end{center}

In this supplementary material, we provide additional details on (I) the derivation of the tripartite Hamiltonian and calculation of the energy spectrum, (II) numerical solution of the master equation, (III) unveiling vacuum fluctuations from Schr\"{o}dinger picture dynamics, and (IV) studying the effect of motional phonon decay. These discussions contribute a comprehensive understanding of our study on unveiling vacuum fluctuations and exploring nonclassical states with cavity-enhanced tripartite interactions.

\section{I.\ \ \ The tripartite Hamiltonian\label{sm_model}}

In this appendix we present the details on the derivation of the tripartite spin-photon-phonon Hamiltonian for the specific laser configurations and level diagram depicted in Fig.~\ref{scheme} of the main text. We focus on a single alkaline-earth-metal atom confined within a single-mode high-finesse cavity with a bare frequency $\omega_c$.  For specificity, the two relevant energy levels for a single $^{88}$Sr atom with one electronic ground state $|g\rangle$ ($^1S_0$) and one long-lived electronic orbital state $|e\rangle$ ($^3P_1$) are included. The atomic narrow (7.5-kHz-wide)  dipole-forbidden transition frequency between $^1 S_0-^3P_1$  is $\omega_e$. This corresponds to the wavelength of $\lambda=689$ nm and the wave vector $k_L=2\pi/\lambda$. As can be seen, the atomic transitions is resonant illuminated by a classical transverse pump field propagating perpendicular to cavity axis with frequency $\omega_p$ and Rabi frequency $\Omega$. To engineer the dynamical spin-orbit coupling, our proposal employsutilizes a single-mode optical cavity to couple the atomic transitions between $|g\rangle\leftrightarrow |e\rangle$, which arises from the collective Bragg scattering. This coupling is characterized by a spatially dependent single atom-cavity coupling $g_0 \sin (k_Lx)$ with $g_0$ being the maximum scattering rate of the single-atom cavity coupling.

Now, considering the level diagram depicted in Fig.~\ref{scheme} of the main text, we can derive the Hamiltonian for the internal states under the rotating-wave approximation (RWA). The Hamiltonian is given by
\begin{align}
\hat{\cal H}_0/\hbar &= \frac{{p}_x^{2}}{2M} +V(x) + {\frac{\Delta}{2}}\sigma_z + \Delta_c \hat{a}^\dag \hat{a} + g_0\sin(k_Lx)(\hat {a}^\dag \hat{\sigma}_{-} + \hat {a}\hat{\sigma}_{+}) +\Omega \hat{\sigma}_{x}, \label{SM-single}
\end{align}
where $M$ is the mass of the atom, $\hat a$ is the annihilation operator of the cavity field, $\sigma_{x,y,z}$ are Pauli matrices representing the spin-$1/2$ system with $\sigma_\pm=(\sigma_x\pm i\sigma_y)/2$, $\Delta_c=\omega_c-\omega_p$ is the cavity-light detuning, and $\Delta=\omega_e-\omega_p$ is the atom-light detuning. In our setup, we consider the single $^{88}$Sr atom confined in a one-dimensional spin-independent harmonic trap $V(x)= \frac{1}{2}M \omega_b^2 x^2$, where $\omega_b$ is the adjustable trap frequency. The trap frequency is respect to the zero-point fluctuation amplitude of the trapped atom oscillator $x_{\rm ZPF}=\sqrt{{\hbar}/{2M\omega_b}}$. In the Lamb-Dicke regime with $k_L x_{\rm ZPF}\ll 1$~\cite{PhysRevA.52.4214}, the Hamiltonian (\ref{SM-single}) can be simplified to
\begin{align}
\hat{\cal H}_1 &= \frac{{p}_x^{2}}{2M} +V(x) + {\frac{\Delta}{2}}\sigma_z + \Delta_c \hat{a}^\dag \hat{a} + g_0 k_Lx (\hat {a}^\dag \hat{\sigma}_{-} + \hat {a}\hat{\sigma}_{+}) +\Omega \hat{\sigma}_{x}. \label{single-LD}
\end{align}

To gain further insight, we can rewrite the single atom-cavity Hamiltonian (\ref{single-LD}) in terms of position-momentum representation
\begin{align}
x=\sqrt{\frac{\hbar}{2M\omega_b}}(\hat b^\dag + \hat
b),~~~ p_x=i\sqrt{\frac{{M\hbar\omega_b}}{2}}(\hat b^\dag - \hat
b),
\end{align}
where $\hat{b}$ is the annihilation operator of bosonic motional phonon for harmonic oscillator frequency $\omega_b$. The operators of motional phonon satisfies the commutation relations, $[\hat{b}, \hat{b}^\dag] =1$. With these operators, we can derive a tripartite spin-photon-phonon Hamiltonian as follows
\begin{align}
\hat{\cal {H}}_2/\hbar &= \omega_b\hat{b}^\dag\hat{b} + \Delta_c \hat{a}^\dag\hat{a} + {\frac{\Delta}{2}}\sigma_z +\Omega \hat{\sigma}_{x} + g (\hat {b}^\dag +\hat {b})(\hat {a}^\dag \hat{\sigma}_{-} + \hat {a}\hat{\sigma}_{+}),
\label{Ham}
\end{align}
where $g=g_0k_Lx_{\rm ZPF}$ is the effective photon-phonon coupling. Apparently, this cavity-enhanced tripartite Hamiltonian $\hat{\cal {H}}$ introduces a novel three-body interactions, which hold significant importance and interest, especially for an essential building block in quantum information processing and photonics community. Notably, this tripartite Hamiltonian enables swapping the excitations among the three distinct degrees of freedom in quantum systems. This capability opens up exciting possibilities for manipulating and exchanging quantum information in integrated quantum systems.

To capture the essence of physics, we introduce the unitary transformation in the rotating frame, defined as
\begin{align}
{\cal U} &= \exp\{-i[\Delta_c'\hat{a}^{\dagger}\hat{a}+ \omega' \hat{b}^{\dagger}\hat{b}]t\},
\end{align}
then tripartite spin-photon-phonon Hamiltonian (\ref{Ham}) reduces to
\begin{align}\label{H2}
{\cal\hat{H}}_3/\hbar &= {\cal U}^{\dagger}{\cal \hat{H}}_2{\cal U} - i{\cal U}^{\dagger}\frac{\partial}{\partial{t}}{\cal U} \nonumber \\
&= \delta_b\hat{b}^\dag\hat{b} + \delta_a \hat{a}^\dag\hat{a} + {\frac{\delta_a-\delta}{2}}\sigma_z +\Omega \hat{\sigma}_{x} + g (\hat {b}^\dag e^{i\omega' t} +\hat {b}e^{-i\omega' t})(\hat {a}^\dag \hat{\sigma}_{-}e^{i\Delta_c' t} + \hat {a}\hat{\sigma}_{+}e^{-i\Delta_c' t} ),
\end{align}
with $\delta_b=\omega-\omega'$, $\delta_a=\Delta_c-\Delta_c'$, and $\delta=\delta_a-\Delta$ being the detuning of the cavity radiation frequency from the atomic resonance.

Under the condition of red-sideband resonance $\omega'=\Delta_c'\approx \omega_b$, the nonrotating wave coupling terms with the high-frequency prefactor $e^{\pm i(\omega'+\Delta_c')t}$ in Eq.~(\ref{H2}) can be neglected using the rotating wave approximation (RWA) with $g/\omega_b\ll 1$. This simplification leads to a `beamsplitter' interactions with respect to the annihilation of a motional phonon up-convert into a quantized cavity photon
\begin{align}\label{BS}
{\cal\hat{H}}_{\rm{B}}/\hbar
&= \delta_b\hat{b}^\dag\hat{b} + \delta_a \hat{a}^\dag\hat{a} +  {\frac{\delta_a-\delta}{2}}\sigma_z +\Omega \hat{\sigma}_{x} + g (\hat {a}^\dag \hat {b} \hat{\sigma}_{-} + \hat {b}^\dag\hat {a}\hat{\sigma}_{+}),
\end{align}
This nonlinear anti-Stokes process represents the coherent exchange of cavity and motional phonon. When the pump field is absent ($\Omega=0$), the Hamiltonian (\ref{BS}) possesses a ${\cal U}(1)$ symmetry characterized by the action of the operator
\begin{align}
{\cal R}_{\theta} =\exp[-i\theta(2\hat{a}^\dag\hat{a}+\hat{b}^\dag\hat{b} +\sigma_z/2)],
\end{align}
which transforms the operators as
\begin{align}
{\cal R}_{\theta}^\dag(\hat{a},\hat{b},\sigma_-
,\sigma_+){\cal R}_{\theta} =(\hat{a} e^{-i2\theta},\hat{b} e^{-i\theta},\sigma_-e^{-i\theta},\sigma_+e^{i\theta}),
\end{align}
The breaking of the ${\cal U}(1)$ symmetry breaking is related to single-atom superradiant phase transitions.

For blue-sideband resonance $\omega'=-\Delta_c'\approx \omega_b$, the rotating wave coupling terms  in Eq.~(\ref{H2}) with the high-frequency prefactor $e^{\pm i(\omega'-\Delta_c')t}$ are safely neglected under RWA with $g/\omega_b\ll 1$. This simplification leads to a `squeeze'  interactionsHamiltonian with respect to the pump photon is down-converted into correlated photon-phonon pairs
\begin{align}\label{Sque}
{\cal\hat{H}}_{\rm{S}}/\hbar
&=  \delta_b\hat{b}^\dag\hat{b} + \delta_a \hat{a}^\dag\hat{a} +  {\frac{\delta_a-\delta}{2}}\sigma_z +\Omega \hat{\sigma}_{x} + g(\hat {a}^\dag \hat {b}^\dag \hat{\sigma}_{-} + \hat {b}\hat {a}\hat{\sigma}_{+}),
\end{align}
This nonlinear Stokes process denotes the amplification and entanglement of the cavity and the motional phonon. In the absence of the pump field ($\Omega=0$), the Hamiltonian (\ref{BS}) possesses another ${\cal U}(1)$ symmetry characterized by the action of the operator
\begin{align}
{\cal R}_{\theta} =\exp[-i\theta(\hat{a}^\dag\hat{a}+\hat{b}^\dag\hat{b} +\sigma_z)]
\end{align}
which yields
\begin{align}
{\cal R}_{\theta}^\dag(\hat{a},\hat{b},\sigma_-
,\sigma_+){\cal R}_{\theta} =(\hat{a} e^{-i\theta},\hat{b} e^{-i\theta},\sigma_-e^{-i2\theta},\sigma_+e^{i2\theta}).
\end{align}
For single-atom superradiance,  strong nonclassical correlated photon-phonon pairs at the single-quanta level are expected, as the emitted photon and phonon are simultaneously created or destroyed. 

Regarding the energy spectrum, the analytical results for the tripartite `squeeze'  interactions can be  obtained by diagonalizing the Hamiltonian ({\ref{Sque}}) for first ignoring the weak classical pump field. It is worth noting that the difference in excitation number between the cavity mode and the motional phonon is $zero$ due to the deterministic parametric down-conversion process. Therefore the restricted Hilbert spaces for the single atom-cavity system are $|n_a-1,n_b-1,e \rangle$ and $|n_a,n_b,g \rangle$, where $n_a$ ($n_b$) denotes the photon (phonon) number excitation of the system. Explicitly, the energy eigenvalues of system with $\delta_a=\delta_b$ fixed, are given by
\begin{align}\label{Energy spectrum}
E_{n_a,n_b,\pm} &=  (n_a+n_b) \delta_{a} - \frac{\delta_a+\delta}{2} \pm \frac{1}{2}\sqrt{4g^{2}n_an_b + (\delta_a+\delta)^{2}},
\end{align}
Assuming $n_a = n_b = n$ without loss of generality, the energy spectrum can be expressed as
\begin{align}\label{Energy spectrum}
E_{n,n,\pm} &=  2n \delta_{a} - \frac{\delta_a+\delta}{2} \pm \frac{1}{2}\sqrt{4g^{2}n^2 + (\delta_a+\delta)^{2}},
\end{align}
corresponding to the single-quanta resonance ($n=1$) 
\begin{align}\label{splitting}
\delta^{(\pm)} _a&= \frac{1}{2}\delta \pm \frac{1}{2}\sqrt{2g^{2} +\delta^2},
\end{align}
where $+$ ($-$) represents the higher (lower) branch. It is noteworthy that the single-quanta resonance $\delta^{(\pm)} _a$ idepends on the value of $\delta$, which is different from the realization of single-photon pairs for bimodal cavities in Ref.~\cite{PhysRevA.105.043711} by utilizing the quantum
interference suppressed two-photon excitation. The detuning $\delta$-enhanced vacuum Rabi splittings between the first pair of dressed states $|1,1,\pm \rangle$  can significantly facilitate the realization of high-quality single-quanta sources beyond the strong coupling regime in experiments. 

\section{II.\ \ \ Numerical solution of the master equation}
To explore the quantum statistics of photon and phonon emissions, we numerically solve the master equation in the dynamics and steady state using the quantum optics toolbox~\cite{tan1999computational}.  In the presence of the complete dissipations of the system, the time evolution of the density matrix $\rho$ dictated by the full Hamiltonian (\ref{triHam}) in the main text obeys the master equation
\begin{equation}
\frac{d{\rho}}{dt} = -i [{\cal {\hat H}}, {\rho}] + \frac{\gamma}{2} \mathcal
{\cal{D}}[{\hat\sigma}_-]\rho + \frac{\kappa_a}{2} \mathcal
{\cal{D}}[\hat{a}]\rho +  \frac{\kappa_b}{2} \mathcal
{\cal{D}}[\hat{a}]\rho,
\label{master}
\end{equation}
where $\gamma$ is atomic spontaneous emission rate of the long-lived excited state, $\kappa_a$ and $\kappa_b$ is, respectively, the decay rate of the photon and phonon, and $\mathcal {D}[\hat{o}]\rho=2\hat{o} {\rho} \hat{o}^\dag - \hat{o}^\dag \hat{o}{\rho} - {\rho} \hat{o}^\dag \hat{o}$ denotes the Lindblad type of dissipation. We should emphasize the standard Lindblad equation (\ref{master}) is applicable for characterizing the strong coupling tripartite spin-photon-phonon interaction~\cite{PhysRevA.84.043832,PhysRevLett.109.193602},  since the Hamiltonian  \eqref{triHam} is essentially originating from RWA with the condition $g/\omega_c\ll1$. We neglect the weak pure dephasing effects in the long-lived atom and phonon fields.

To characterize  the nonclassical correlations, we utilize the standard second-order correlation function~\cite{PhysRev.130.2529}
\begin{align}
g_{oo}^{(2)}(\tau)=\frac{\langle\hat{o}^\dagger(t)\hat{o}^\dagger(t+\tau)
\hat{o}(t+\tau)\hat{o}(t)\rangle}{\langle\hat{o}^\dagger(t)
\hat{o}(t)\rangle^2},
\end{align}
with $o=a, b$. For $\tau=0$,  $g_{aa}^{(2)}(0)$ [$g_{bb}^{(2)}(0)$] represents the cavity (phonon) mode and is a key physical quantity for assessing the purity of single-quanta emissions~\cite{senellart2017high}.  The value of $g_{oo}^{(2)}(0)$ should be less than 1 to ensure sub-Poissonian statistics. Furthermore, for nonzero time interval, $g_{oo}^{(2)}(\tau)$ can be straightforwardly calculated using the quantum regression theorem. Experimental measurements of $g_{oo}^{(2)}(\tau)$ can be measured using techniques such as Hanbury, Brown, and Twiss interferometer for cavity photons~\cite{birnbaum2005photon} and spin state-resolved projective measurements for motional phonons~\cite{PhysRevX.8.021027,wolf2019motional,PhysRevLett.128.183601}.  Now, the quantum statistics for PB should satisfy two conditions: $g_{oo}^{(2)}(0)<1$ to ensure the sub-Poissonian statistics and $g_{oo}^{(2)}(0)<g_{oo}^{(2)}(\tau)$ to ensure antibunching with respect to single photon or phonon.

\section{III.\ \ \ Resolving Vacuum Fluctuations from Schr\"{o}dinger picture dynamics}

In order to gain further insight, we simplify the analytical model by neglecting system damping and the weak pump field. This allows us to resolve the vacuum fluctuations of the atom oscillator's motional phonon and the quantized cavity field from the Schr\"{o}dinger picture dynamics.  We consider both the photon and phonon fields to be in Fock states $|n,n\rangle$, while the single atom is in the excited state $|e\rangle|$.

\subsection{The tripartite `squeeze'  interactions}
For the resonance case with $\delta_a=\delta_b=\delta=0$, the tripartite `squeeze' Hamiltonian simplifies to
\begin{align}\label{SMtriSque}
{\cal\hat{H}}_{\rm{S}}/\hbar
&=  g(\hat {a}^\dag \hat {b}^\dag \hat{\sigma}_{-} + \hat {b}\hat {a}\hat{\sigma}_{+}).
\end{align}
It is worth noting that the difference in excitation number between the photon and phonon fields is zero due to the deterministic parametric down-conversion process. As a result, the restricted Hilbert spaces for the single atom-cavity system
are $|n,n,e\rangle$ and $|n+1,n+1,g\rangle$. The energy eigenvalues of Eq.~(\ref{SMtriSque}) are given by
\begin{align}
\lambda_{\pm}(n,n)/\hbar = \pm g(n+1),
\end{align}
and the corresponding eigenstates $|n,\pm\rangle$ associated with these energy eigenvalues satisfy
\begin{align}
|n,n,+\rangle &= [|n,n,e\rangle + |n+1,n+1,g\rangle]/\sqrt{2} \nonumber \\
|n,n,-\rangle &= [|n,n,e\rangle - |n+1,n+1,g\rangle]/\sqrt{2}.
\end{align}

It is now possible to obtain the dynamics of a general state by expanding it on to the dressed eigenstates. We consider an initial state for the quantized fields that satisfies $|\psi_{{\rm {field}}}(t=0)= |n,n\rangle$ and assume the single atom is injected in the long-lived excited state. Thus the initial state for the tripartite system reads
\begin{align}
|\psi(t=0)\rangle &= [|n,n,+\rangle + |n,n,-\rangle]/\sqrt{2},
\end{align}
Since $|n,\pm \rangle$ are stationary states of the tripartite spin-photon-phonon system, the state vector with time development can be straightforwardly calculated as follows
\begin{align}
|\psi(t)\rangle &=e^{-i{\cal\hat{H}}_{\rm{S}}/\hbar}|\psi(t=0)\rangle =  [|n,n,+\rangle e^{-i\lambda_{+}(n)t/\hbar}+ |n,n,-\rangle e^{-i\lambda_{-}(n)t/\hbar}]/\sqrt{2}, \nonumber \\
&=\cos\theta_S|n,n,e\rangle -i \sin\theta_S |n+1,n+1,g\rangle, 
\end{align}
where $\theta_S=g(n+1)t$ represents the accumulated phases after the tripartite `squeeze'  interactions. It is clear that the state vector exhibits Rabi oscillations. Furthermore, if both the photon and motional phonon are initially in vacuum states, the state of the tripartite system as a function of time can be expressed as
\begin{align}
|\psi(t)\rangle  =  \cos(gt)|0,0,e\rangle -i \sin(gt) |1,1,g\rangle.
\end{align}
Clearly, the system undergoes Rabi oscillations between the vacuum state and a state with one photon and one phonon.

\subsection{The tripartite `beamsplitter'  interactions}

For the resonance case with $\delta_a=\delta_b=\delta=0$, the tripartite `beamsplitter' Hamiltonian reduce to
\begin{align}\label{SMtriBS}
{\cal\hat{H}}_{\rm{B}}/\hbar
&={\frac{\Delta}{2}}\sigma_z +\Omega \hat{\sigma}_{x} + g (\hat {a}^\dag \hat {b} \hat{\sigma}_{-} + \hat {b}^\dag\hat {a}\hat{\sigma}_{+}),
\end{align}
assuming that the field modes are initially in Fock states $|n,n\rangle$. Therefore the restricted Hilbert spaces for the single atom-cavity system are $|n,n,e\rangle$ and $|n+1,n-1,g\rangle$. The energy eigenvalues of Eq.~(\ref{SMtriBS}) are given by\begin{align}
\lambda_{\pm}(n,n)/\hbar = \pm g\sqrt{n(n+1)},
\end{align}
and the eigenstates $|n,\pm\rangle$ associated with the energy eigenvalues satisfy
\begin{align}
|n,n,+\rangle &= [|n,n,e\rangle + |n+1,n-1,g\rangle]/\sqrt{2} \nonumber \\
|n,n,-\rangle &= [|n,n,e\rangle - |n+1,n-1,g\rangle]/\sqrt{2}.
\end{align}

It is now possible to obtain the dynamics of a general state by expanding it on to the dressed eigenstates. We consider an initial state for the quantized fields that satisfies $|\psi_{{\rm {field}}}(t=0)= |n,n\rangle$ and assume the single atom is injected in the long-lived excited state. Thus the initial state for the tripartite system reads
\begin{align}
|\psi(t=0)\rangle &= [|n,n,+\rangle + |n,n,-\rangle]/\sqrt{2},
\end{align}
Since $|n,\pm \rangle$ are stationary states of the tripartite spin-photon-phonon system, the state vector with time development can be straightforwardly calculated as follows
\begin{align}
|\psi(t)\rangle &=e^{-i{\cal\hat{H}}_{\rm{B}}/\hbar}|\psi(t=0)\rangle =  [|n,n,+\rangle e^{-i\lambda_{+}(n,n)t/\hbar}+ |n,n,-\rangle e^{-i\lambda_{-}(n,n)t/\hbar}]/\sqrt{2}, \nonumber \\
&= \cos\theta_B |n,n,e\rangle -i \sin\theta_B |n+1,n-1,g\rangle,
\end{align}
associated with $\theta=g\sqrt{n(n+1)}t$ ($n>0$) being the accumulated phases after tripartite `beamsplitter'  interactions. Interestingly, consider both photon and motional phonon are initially in the vacuum states, the state of the tripartite system as a function of time satisfies $|\psi(t)\rangle  =|0,0,e\rangle$, which corresponds to zero average photon and phonon  occupancies.

\section{IV.\ \ \ The effect of motional phonon decay}

\begin{figure}[ht]
\includegraphics[width=0.7\columnwidth]{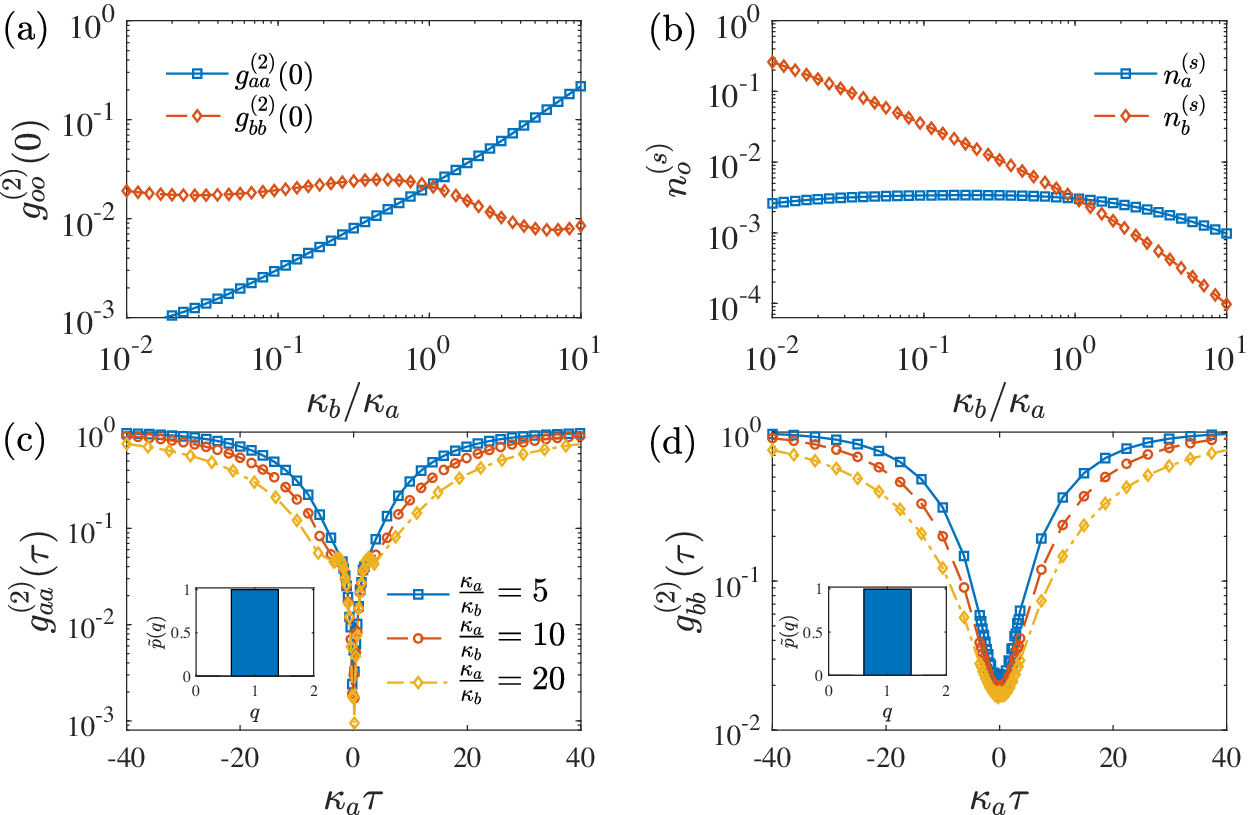}
\caption{(color online). (a) and (b) shows the phonon decay $\kappa_b$ dependence of $g^{(2)}_{oo}(0)$ and $n^{(s)}_{o}$, respectively. (c)-(d) Time interval $\tau$ dependence of $g^{(2)}_{aa}(\tau)$ and $g^{(2)}_{bb}(\tau)$ for different values of $\kappa_b$. Inset shows the steady-state distribution $\tilde p(q)$  with $\kappa_a/\kappa_b =10$. The other parameters are $\delta_a=\delta_b=g(1+\sqrt{3})/2$, $\delta/g=1$, and $\omega'/g=100$.} \label{g2time}
\end{figure}

In this section, we study the effect of motional phonon decay on the quantum coherence and emission properties of the tripartite spin-photon-phonon system. We numerically calculate the second-order correlation function $g^{(2)}_{oo}(0)$ and the corresponding steady-state population $n^{(s)}_{o}$ with leaving the motional phonon decay rate $\kappa_b$ as a tunable parameter. The results are shown in Figs.~\ref{g2time}(a) and \ref{g2time}(b). We observe that a larger phonon decay rate leads to a stronger antibunching amplitude ($g^{(2)}{aa}(0)$) and a smaller occupation number of emissions ($n^{(s)}{a}$). It is clear that the decay rate of the motional phonon significantly affects both the photon and phonon emissions. As $\kappa_b$ increasing, the antibunching amplitude $g^{(2)}_{aa}(0)$ rapidly grow, albeit the photon excitation $n^{(s)}_{a}$ remains roughly unchanged when $\kappa_b/\kappa_a<1$. As for motional phonon, the value of $g^{(2)}_{bb}(0)$  is insensitive to variations in $\kappa_b$, but the phonon excitation $n^{(s)}_{b}$ rapidly decreases with increasing $\kappa_b$. Compared to the long-lived motional phonon, the antibunching amplitude for photon is significantly enhanced by over one orders of magnitude with up to $g^{(2)}_{aa}(\tau)<10^{-3}$ when the decay rate ratio satisfies $\kappa_b/\kappa_a\sim 0.01$. Of importance, the proposed tripartite system exhibiting both strong photon- and phonon-blockade and corresponding to a large population of steady state can be used as a high-quality quantum sources.

To further analyze the single-quanta blockade nature of photon and phonon emissions, we plot the interval $\tau$ dependence of the second-order correlation functions $g^{(2)}_{aa}(\tau)$ (c) and $g^{(2)}_{bb}(\tau)$ for different values of  $\kappa_b$, as shown in Figs.~\ref{g2time}(c) and \ref{g2time}(d). Interestingly, the decay times for both $g^{(2)}_{aa}(\tau)$ (c) and $g^{(2)}_{bb}(\tau)$ are obviously longer than the typical time scale of $\kappa_a^{-1}$. This behavior is a result of the advantages of the long-lived motional phonon in the tripartite spin-photon-phonon system. We show that the decay of antibunching with $g^{(2)}_{oo}(\tau)>g^{(2)}_{oo}(0)$ decreases gradually as the decay ratio $\kappa_a/\kappa_b$ increases. 

To characterize the single-quanta blockade nature of the photon and phonon emissions, we examine the steady-state distribution $\tilde p(q)\equiv qp(q)/n^{(s)}$ which represents the fraction of $q$-quanta states among the total excitations of the steady state. Clearly, the strong photon and phonon blockade with the typical probability of multiquanta excitations ($n>2$) being below $0.001\%$ and $0.07\%$ are achieved, respectively, for $\kappa_a/\kappa_b=10$. These results demonstrate that the proposed tripartite system can serve as a high-quality quantum source with strong photon and phonon blockade. 

\end{document}